\documentclass[12pt,a4paper]{article}
\usepackage{amsmath}
\usepackage{amsfonts}
\usepackage{amssymb}
\usepackage{textcomp}
\usepackage[dvips]{graphicx}
\usepackage{epsfig}
\usepackage{bm}
\usepackage{dcolumn}
\usepackage{amsfonts}
\usepackage{color}

\usepackage[left=2.27cm,right=2.27cm,top=2.27cm,bottom=2.72cm]{geometry}
\begin{document}
\title{\textbf{Slow-roll approximations in Einstein--Gauss--Bonnet gravity formulated in terms of e-folding numbers}}
\author{E.O.~Pozdeeva$^{a,}$\footnote{E-mail: pozdeeva@www-hep.sinp.msu.ru},  \\
\small$^a$ Skobeltsyn Institute of Nuclear Physics, Lomonosov Moscow State University,\\ \small  Leninskie Gory~1, Moscow 119991, Russia}
\date{ \ }
\maketitle
\begin{abstract}
In the Einstein--Gauss--Bonnet (EGB) gravity models, the slow-roll approximation has been extended by taking into account the first-order slow-roll parameter $\delta_1 =-2\,H^2\,\xi^\prime/U_0$,  which is proportional to the first derivative of the Gauss-Bonnet coupling function $\xi$ with respect to the e-folding number. These extensions lead to the question of the accuracy of effective potential reconstruction during the generalization of attractors in EGB gravity. We have reconstructed models using the extended slow-roll approximations and compared them with the exact expressions and the standard slow-roll approximation.
\end{abstract}

{PACS: 98.80.-k, 98.80.Cq, 04.50.Kd}


\section*{Introduction}\label{introduction}
The models of EGB gravity are interesting for the current study of various aspects of the evolution of the Universe  \cite{Yogesh:2024zwi,Oikonomou:2024jqv,Mudrunka:2023wxy,Odintsov:2023lbb,Nojiri:2023mvi,Odintsov:2023aaw,Kawaguchi:2022nku,Odintsov:2022zrj,Gangopadhyay:2022vgh,Oikonomou:2022ksx,
Younesizadeh:2021heg,Oikonomou:2021kql,Kawai:2021edk,Kawai:2021bye,Pozdeeva:2021nmz,Rashidi:2020wwg,Kleidis:2019ywv,Odintsov:2019clh,
Odintsov:2018zhw,Koh:2018qcy,Yi:2018gse,Chakraborty:2018scm,Koh:2016abf,vandeBruck:2016xvt,Mathew:2016anx,Hikmawan:2015rze,Koh:2014bka,
Jiang:2013gza,Guo:2009uk,Tsujikawa:2006ph,Pozdeeva:2020apf,Torii:1996yi}.

In Ref. \cite{Pozdeeva:2020shl}, the slow-roll approximation was used to obtain appropriate models in EGB gravity, leading to models with inflationary parameters such as in the attractors models \cite{Galante:2014ifa}. As a result, a set of models with the Hubble parameter proportional to an exponential function of inverse  number of e-foldings and the Gauss-Bonnet coupling function inversely proportional to the Hubble parameter were obtained. These models were tested in accordance with known models \cite{Mukhanov:2013tua,Bezrukov:2007ep}. In Ref. \cite{Pozdeeva:2021iwc}, we extended the expression for the spectral index by including the second order of the inverse e-folding number in EGB gravity for generalized attractors models. We used the extended expression for the spectral index in terms of e-folding numbers to estimate the beginning of inflation. To generalize the attractor behavior to EGB gravity and reconstruct the form of potential the standard slow-roll approximation was used.

In Ref. \cite{Pozdeeva:2024ihc}, the question of accuracy of the slow-roll approximation for EGB models with monomial potentials was considered.  Two new slow-roll approximations extended by the first order of parameter $\delta_1$ were obtained during the study.  In \cite{Pozdeeva:2024ihc} the numeral simulations indicated the insufficient accuracy of the standard slow-roll approximation to study the evolution of EGB gravity models with a monomial potential. These new slow-roll approximations have been applied to EGB models with potential $V\sim\left(\frac{\phi^n+c}{\phi^n}\right)$ in Ref. \cite{Yogesh:2024mpa}.

We compare the standard slow-roll approximation with the extended slow-roll approximations taking into account correction of the first order and the exact expression for potential. The model potential in the standard slow-roll approximation was suggested in \cite{Pozdeeva:2020shl} and it was applied to generate the set of models in EGB gravity with a scalar field non-minimally coupled with gravity \cite{Pozdeeva:2021iwc} due to the effective potential \cite{Pozdeeva:2019agu}.

In this paper, we study the applicability of two new extended slow-roll approximations and the standard slow-roll approximation to the model with the Hubble parameter
which is an exponential function from inverse e-folding number and the Gauss--Bonnet coupling inversely proportional to the squared Hubble parameter. In Section \ref{model}, we introduce a general model of EGB gravity and define three types of slow-roll approximations in the most general form. In Section \ref{ExpModel}, we present the basis of the exponential models and apply it to the formulas from Section \ref{model}.
In Section \ref{NumEst}, we  discuss the choice of model parameters in detail, and study the deviation of the slow-roll approximations from exact behavior numerically for the natural choice of parameters. In Section \ref{Conclusion}, we summarize the results and evaluate the applicability of these approximations.
\section{Slow-roll approximations in EGB gravity}\label{model}
In this paper, we consider the slow-roll approximations for models with the Gauss--Bonnet term $\mathcal{G}=R_{\mu\nu\rho\sigma}R^{\mu\nu\rho\sigma}-4R_{\mu\nu}R^{\mu\nu}+R^2\,$,  described by the following action:
\begin{equation}
\label{action1}
S=\int d^4x\sqrt{-g}\left[U_0R-\frac{1}{2}g^{\mu\nu}\partial_\mu\phi\partial_\nu\phi-V(\phi)-\frac{1}{2}\xi(\phi){\cal G}\right],
\end{equation}
where $U_0>0$ is a constant, the functions $V(\phi)$ and $\xi(\phi)$ are differentiable ones, $R$ is the Ricci scalar\footnote{The spatially flat Friedmann  metric with $ds^2=-dt^2+a^2(dx^2+dy^2+dz^2)$ is considered as background.}.

We introduce EGB model and its approximations in terms of e-folding number supposing $\dot{A}=\frac{dA}{dt}=-H\frac{d A}{d N}$ (where $A$ is any function) such as in  \cite{Pozdeeva:2020shl} and present the first slow-roll parameters \cite{Guo:2010jr,vandeBruck:2015gjd} in terms of e-folding numbers \footnote{$H=-\frac{dN}{dt}$, $N=-\ln(a/a_0)$, $a_0$ is any number} :
\begin{eqnarray}
  \epsilon_1 &=&\frac12\frac{d\ln(Q)}{dN},\,\,\delta_1 = -\frac{2\,Q}{U_0}\frac{d\xi}{dN}\label{d1},\,\mbox{where}\quad Q=H^2.\label{ed1}
\end{eqnarray}
The exact expression for the potential can be obtained from representation of the Friedmann equations in terms of derivatives with respect to e-folding number.
 The equations of the model can be formulated such as:
\begin{eqnarray}
  & & 12\,Q\,\left(U_0+2\xi^\prime\,Q\right)=Q\Phi+2V,\label{leads to V} \\
  & & 2\,(Q)^\prime\left(U_0+2\xi^\prime\,Q\right)=Q\Phi-4Q\left(Q\xi^{\prime\prime}+\left(\frac{(Q)^\prime}{2}+Q\right)\xi^\prime\right),\label{subs Phi}
\end{eqnarray}
where ${}^\prime=\frac{d}{dN}$, $\Phi=\chi^2$, $\chi=\frac{d\phi}{dN}$.\\
We get expression for $\Phi$ using \eqref{subs Phi}:
 \begin{equation}\label{Phi}
  \Phi_{exact}=\frac{2\,U_0\,Q^\prime+6\,Q\,Q^\prime\,\xi^\prime+4\,Q^2\left(\xi^{\prime\prime}+\xi^\prime\right)}{Q}
 \end{equation}
 and substitute \eqref{Phi} to \eqref{leads to V} to get the potential $V$:
\begin{equation}\label{Vexact}
    V_{exact}=-\left(3\,{\xi}^\prime\,Q+{U_0}\right)\,{Q}^\prime+ 2\,\left(5\,{\xi}^\prime -{\xi}^{\prime\prime}\right) {Q}^{2}+6\,{U_0}\,Q.
\end{equation}
Varying of action \eqref{action1} with respect to field in the spatially flat Friedmann Universe and reformulating obtained equation in terms of e-folding derivative, we get:
\begin{equation}\label{VarField}
    \frac{Q}{2}\Phi^\prime+\frac{Q^\prime}{2}\Phi-3\,Q\,\Phi=-V^\prime-12\,Q\,\xi^\prime\left(-\frac{Q^\prime}{2}+Q\right).
\end{equation}
In the slow-roll regime ($\ddot{\phi}\ll3H\dot{\phi}$ or equivalently $ \ddot{\phi}\phi^\prime=\frac{Q}{2}\Phi^\prime+\frac{Q^\prime}{2}\Phi\ll3\,Q\,\Phi$) the
equation \eqref{VarField} can be reduced to
\begin{equation}\label{VarFieldReduced}
    3\,Q\,\Phi=V^\prime+12\,Q\,\xi^\prime\left(-\frac{Q^\prime}{2}+Q\right)
\end{equation}
leading to
\begin{equation}\label{ExtraSlowRollVarField}
 \Phi_{1,2}=\frac{V^\prime_{1,2}+12\,Q\,\xi^\prime\left(-\frac{Q^\prime}{2}+Q\right)}{3Q}
\end{equation}
for the extended slow-roll approximations and
 \begin{equation}\label{SlowRollVarField}
 \Phi_{sl}=\frac{V^\prime_{sl}+12\,Q^2\,\xi^\prime}{3Q}
\end{equation}
for the standard slow-roll approximation.

The standard slow-roll approximation supposes  \cite{Pozdeeva:2020shl}:
 \begin{equation}\label{Vsl}
 V_{sl}=6\,U_0\,Q.
 \end{equation}
The expressions of the potential for the expanded slow-roll approximations have been obtained in \cite{Pozdeeva:2024ihc} (see Appendix \ref{potential}):
\begin{eqnarray}
V_1&\approx&\frac{6\,U_0\,Q}{1+\delta_1},\label{V1}\\
V_2&\approx&6\,U_0\,(1-\delta_1)\,Q.\label{V2}
\end{eqnarray}
Substituting \eqref{V2} to  $\Phi_2$  from \eqref{ExtraSlowRollVarField} and using the definition of the slow-roll parameter $\delta_1$ in terms of e-folding number derivatives \eqref{ed1},  we get \eqref{Phi}, thus $\Phi_2=\Phi_{exact}$.
\section{Exponential model in different types of slow-roll approximation}\label{ExpModel}
In this study, we explore the corrections to the slow-roll regime. We begin with a model based on the Hubble parameter and the Gauss-Bonnet coupling function:
\begin{eqnarray}
  Q&=& Q_0\exp\left(-\frac{3}{2}\frac{C_\beta}{(N+N_0)}\right),\\
  \xi &=& \frac{\xi_0\,Q_0}{Q}.
\end{eqnarray}
where $C_\beta$, $N_0$, $Q_0$, $\xi_0$ are model constants. \\
We substitute these expressions into the formulas for the first slow-roll parameters \eqref{d1}:
\begin{eqnarray}
  \epsilon_1 &=&\frac{3}{4}\frac{C_\beta}{(N+N_0)^2}, \\
  \delta_1 &=&-\frac{2\xi_0\,Q_0}{U_0\,Q}\frac{d\,Q}{dN}=\frac{3\,\xi_0\,Q_0}{U_0}\frac{C_\beta}{(N+N_0)^2}.\label{delta1}
\end{eqnarray}
To get exit from inflation at $N=0$ ($\epsilon_1=1$) we put $C_\beta=\frac{4\,N_0^2}{3}$. We obtain following expressions for the inflationary parameters: the tensor-to-scalar ratio, the spectral index of scalar perturbation and the amplitude of scalar perturbation  \cite{vandeBruck:2015gjd,Guo:2010jr,Hwang:2005hb}:
\begin{eqnarray}
  r &\approx& 8|2\epsilon_1-\delta_1|\approx\frac{16\,N_0^2}{(N+N_0)^2}\left|1-\frac{4\,\xi_0\,Q_0}{U_0}\right| ,\label{rs}\\
  \label{ns}n_s&\approx&1-2\epsilon_1+\frac{d\ln(r)}{dN}\approx\,1-\frac{2}{N+N_0}\left(1+\frac{N_0}{N+N_0}\right),\\
  A_s &\approx&\frac{Q}{\pi^2\,U_0\,r} \approx\frac{Q_0(N+N_0)^2\exp\left(-\frac{2\,N_0^2}{(N+N_0)}\right)}{16\pi^2\,U_0\,N_0^2\,\left|1-\frac{4\,\xi_0\,Q_0}{U_0}\right|}.
\end{eqnarray}

These inflationary parameters are mostly interesting at the beginning of inflation for testing with observations \cite{Planck:2018jri,BICEP:2021xfz,Galloni:2022mok}.  The speed of gravitational wave are considered for EGB gravity of inflation in \cite{Oikonomou:2024etl}. The estimation of speed of gravitational waves from observations are presented in \cite{LIGOScientific:2017vwq,LIGOScientific:2017zic}. We study the speed of gravitational waves corresponding to the considering model in Subsection \ref{Wave speeds}. In Subsection \ref{Specter of tensor perturbations}, we estimate the spectral index of tensor perturbations.

We derive the expressions for the potential of the exponential model under consideration using approximations and the exact formula:
\begin{eqnarray}
  & &V_{sl}=6{U_0}{Q_0}\,{\exp\left(-{\frac{2{{N_0}}^{2}}{N+{N_0}}}\right)},\qquad\qquad\qquad\\
  & &V_1={V_{sl}}\left( 1+{\frac {4{Q_0}\,{\xi_0}\,{{N_0}}^{2}}{{U_0}
\, \left( N+{N_0} \right)^{2}}}\right)^{-1},\qquad\\
  & &V_2={V_{sl}} \left( 1-{\frac{4\,{Q_0}\,{\xi_0}\,{{N_0}}^{2}}{{U_0}\, \left( N+{N_0}\right)^{2}}}\right),\qquad\\
& &V_{exact}=V_{sl}\left(1-\frac{N_0^{2}}{3\,(N+{N_0})^{2}}-
\frac{2{Q_0}{\xi_0}{N_0}^{2}(5\,(N+N_0)^2-N_0^{2}+2\,(N+N_0))}{3\,U_0 \,(N+N_0)^{4}}\right).\end{eqnarray}
We substitute the potentials to corresponding expressions for $\Phi=(\phi^\prime)^2$ and get:
\begin{eqnarray}
  & &\Phi_{sl}={\frac {4\,{{N_0}}^{2}\left(U_0-2\,{Q_0}\,{\xi_0}\right)}{ \left( N+{N_0} \right)^{2}}}, \\
& &\Phi_2=\frac{4\,N_0^2}{(N+N_0)^2}\left(U_0-2\,Q_0\xi_0+\frac{4\,Q_0\,\xi_0}{(N+N_0)}-\frac{2\,Q_0\xi_0\,N_0^2}{(N+N_0)^2}\right).
\end{eqnarray}
The field $\phi$ can be expressed analytically for the exponential model. However, to avoid complications during subsequent numerical calculations,
it is better to integrate $\sqrt{\Phi_{exact}}$ after selecting the model parameters. From $\Phi_{exact}=\Phi_{2}$ we get equality of fields $\phi=\phi_{exact}=\phi_2$ for both the exact expression and the approximation \eqref{V2}.

At the same time, the expression  $\Phi_1$ is very long
$$\Phi_1=\frac{4\,U_0^2\,N_0^2\left(1+\frac{4\xi_0\,Q_0(N+N_0)}{U_0(N+N_0)^2+4\xi_0\,Q_0\,N_0^2}\right)}{U_0(N+N_0)^2+4\xi_0\,Q_0\,N_0^2}+\frac{8\,Q_0\xi_0\,N_0^2\left(\frac{N_0^2}{(N+N_0)^2}-1\right)}{(N+N_0)^2}$$
and does not lead to an analytical expression for $\phi_1(N)$ for the considering exponential model.

In the standard slow-roll approximation, the expression for the field's dependence on the e-folding number during inflation has a simple form
\begin{equation}\label{phi_sl}
\phi_{sl}=2\,{N_0}\,\sqrt {U_0-2\,{Q_0}\,{\xi_0}}\ln\left( N+N_0\right)+c_{sl},
\end{equation}
where $c_{sl}$ is constant, which should be chosen in order to compare the behavior of $\phi$ and $\phi_{sl}$ in numerical analysis.
\section{Choice of model parameters, numerical estimation}\label{NumEst}
To analyze the behavior of $\phi(N)$ we should choose the model parameters. Following \cite{Pozdeeva:2021iwc} and the cosmological attractors models
\cite{Galante:2014ifa}, we write \eqref{rs} as $r=\frac{12\,C_\alpha}{(N_b+N_0)^2}$ and get:
\begin{equation}\label{Attr}
    N_0^2\left|1-\frac{4\xi_0\,Q_0}{U_0}\right|=\frac34\,C_\alpha,
\end{equation}
where $C_\alpha$ is a constant restricted by observations \cite{Planck:2018jri,BICEP:2021xfz,Galloni:2022mok}.\\
At the choice \eqref{Attr}, the expression for the scalar perturbation amplitude at
the beginning of inflation $(N=N_b)$ is
\begin{equation}\label{As}
    A_s|_{N=N_b}=\frac{Q_0\,(N_b+N_0)^2\,\exp\left(-\frac{2\,N_0^2}{N_b+N_0}\right)}{12\pi^2\,U_0^2\,C_\alpha},\end{equation}
from where we can determine the value of $Q_0$:
\begin{equation}\label{H02}
    Q_0=\frac{12\pi^2\,U_0^2\,C_\alpha\cdot\left(A_s|_{N=N_b}\right)\cdot\exp\left(\frac{2\,N_0^2}{N_b+N_0}\right)}{(N_b+N_0)^2}.
\end{equation}
Fixing $N_0^2$, $C_\alpha$ and ${Q_0}$ we automatically fix $\xi_0$, such as
\begin{equation}\label{modulScobka}
\left|1-\frac{4\xi_0{Q_0}}{U_0}\right|=\frac34\frac{C_\alpha}{N_0^2}.
 \end{equation}
The expression in the module brackets can be positive or negative:
\begin{itemize}
  \item if $\frac{4\xi_0{Q_0}}{U_0}<1,$ then $\xi_0=\xi_{01}=\frac{U_0}{4Q_0}\left(1-\frac{3}{4}\frac{C_\alpha}{N_0^2}\right)$
  from where
  \begin{enumerate}
    \item if $\frac34\,C_\alpha<N_0^2$, then $\xi_0$ is positive,
    \item if $N_0^2<\frac34\,C_\alpha$, then $\xi_0$ is negative;
  \end{enumerate} which is possible for sufficiently large positive $\xi_0$, then
  \item if $1<\frac{4\xi_0{Q_0}}{U_0},$ when $\xi_0=\xi_{02}=\frac{U_0}{4\,Q_0}\left(\frac{3}{4}\frac{C_\alpha}{N_0^2}+1\right),$ which corresponds to positive $\xi_0$.
\end{itemize}
Now we start numerical estimations. We define the constant coupling  $U_0$  as  $U_0={M_{Pl}^2}/{2}$. For convenience,
we put $N_0=1$. For simplicity, we follow to estimation for tensor-to-scalar ratio $r$ corresponding to $R+R^2$ inflationary model \cite{Starobinsky:1980te} using $C_\alpha=1$.

Using the observations of the cosmic microwaves background \footnote{$A_s=(2.10\pm 0.03)\times 10^{-9},$ $n_s=0.9654\pm 0.0040,$ $r < 0.028$ } \cite{Planck:2018jri,BICEP:2021xfz,Galloni:2022mok}, we introduce values of the model parameters.
We estimate the starting point of inflation, $N=N_b$, by substituting the spectral index value $n_s=0.9654$ into equation \eqref{ns}:
\begin{equation}\label{Nb}
    N_b\approx57.787.
\end{equation}
We apply expression for $Q_0$ \eqref{H02} by substituting $A_s=2.1\cdot 10^{-9}$, and obtain:
\begin{equation}
Q_0\approx1.8861\cdot10^{-12}\pi^2.\label{Q0}
\end{equation}
We calculate $\xi_0=\xi_{01}$ assuming that the expression in the module brackets is positive:
\begin{equation}\xi_0\approx1.6788\cdot10^{10}/\pi^2.\label{xi0}\end{equation}
For obtained model parameters, we get $r\approx 0.0035$ at the beginning of inflation.\\
In \cite{Pozdeeva:2019agu}, we introduce the effective potential for EGB gravity. The effective potential $V_{eff}$ and it's equivalent presentations  play am important role in the   study of de Sitter solutions stability \cite{Pozdeeva:2019agu,Vernov:2021hxo}. Here, we consider an equivalent presentation of the effective potential $\tilde{V}_{eff}$:
\begin{equation}
\tilde{V}_{eff}=-V_{eff}^{-1},\quad  \mbox{where}\quad  V_{eff}=-U_0^2/V+\xi/3.
\end{equation}
Here we should note, that the considering set of the model parameters is mostly toy and used to clear present the behavior of effective potential $\tilde{V}_{eff}$ which is analog to potential in the General Relativity models.

\begin{figure}[htp]
\includegraphics[scale=0.25]{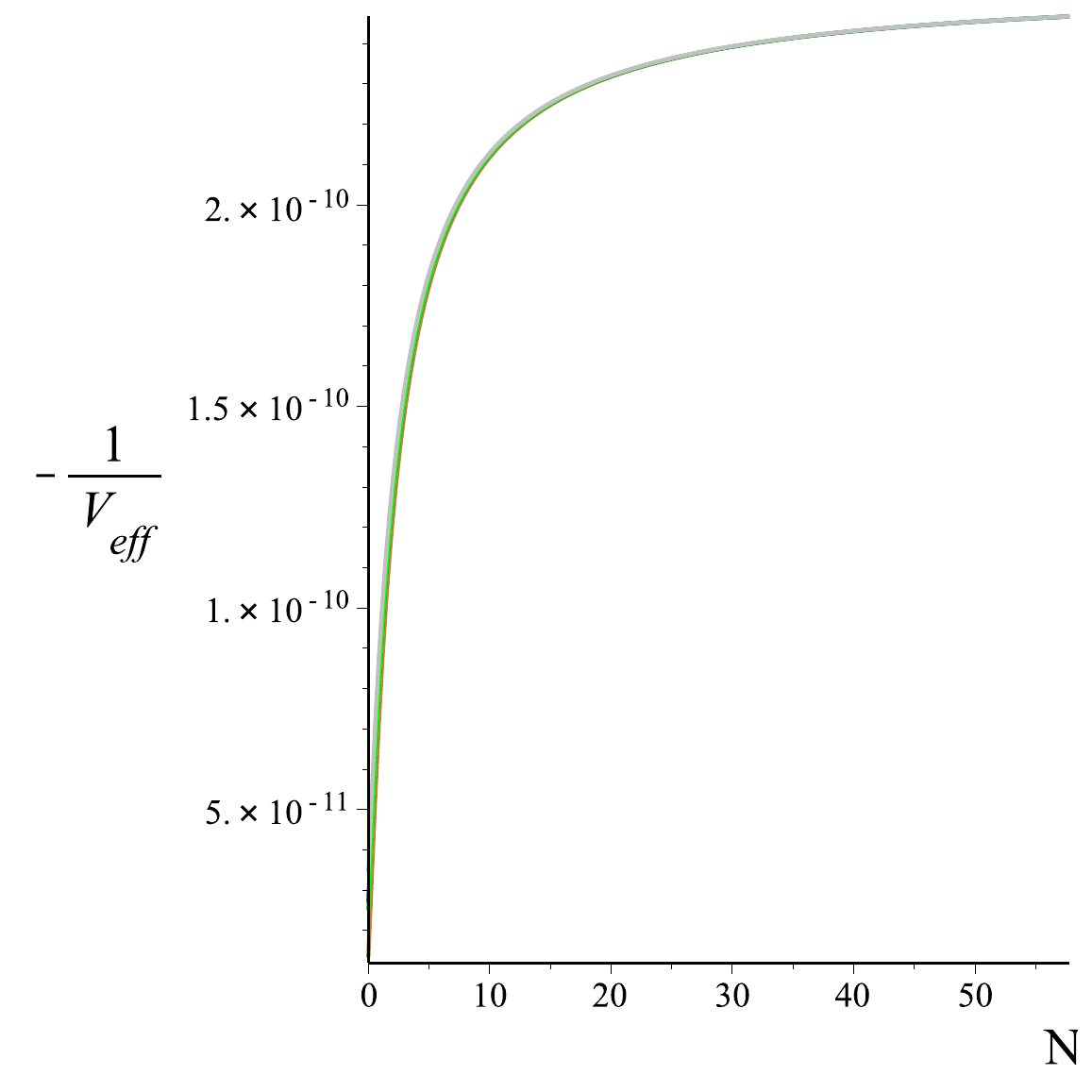}\quad
\includegraphics[scale=0.25]{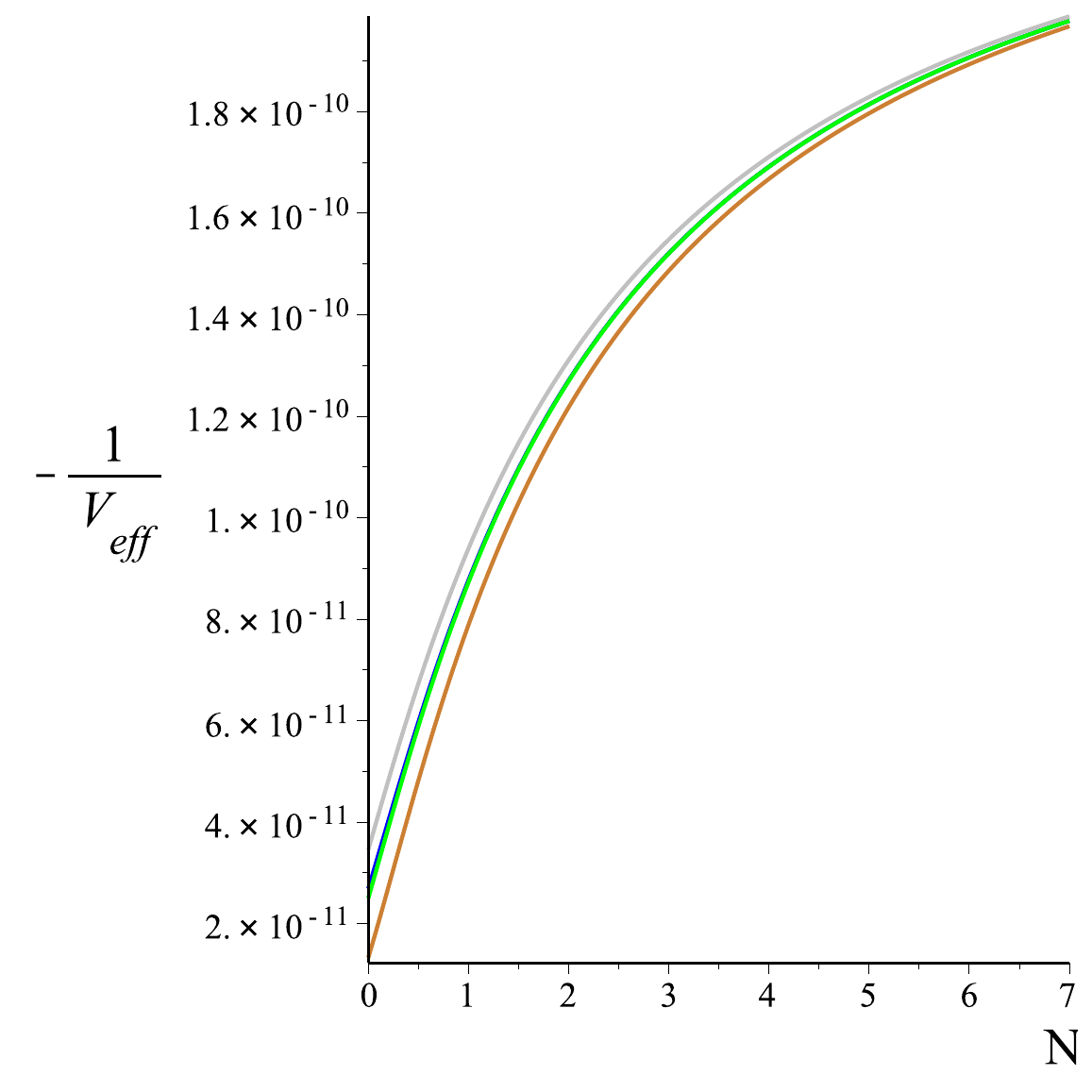}\quad
\includegraphics[scale=0.25]{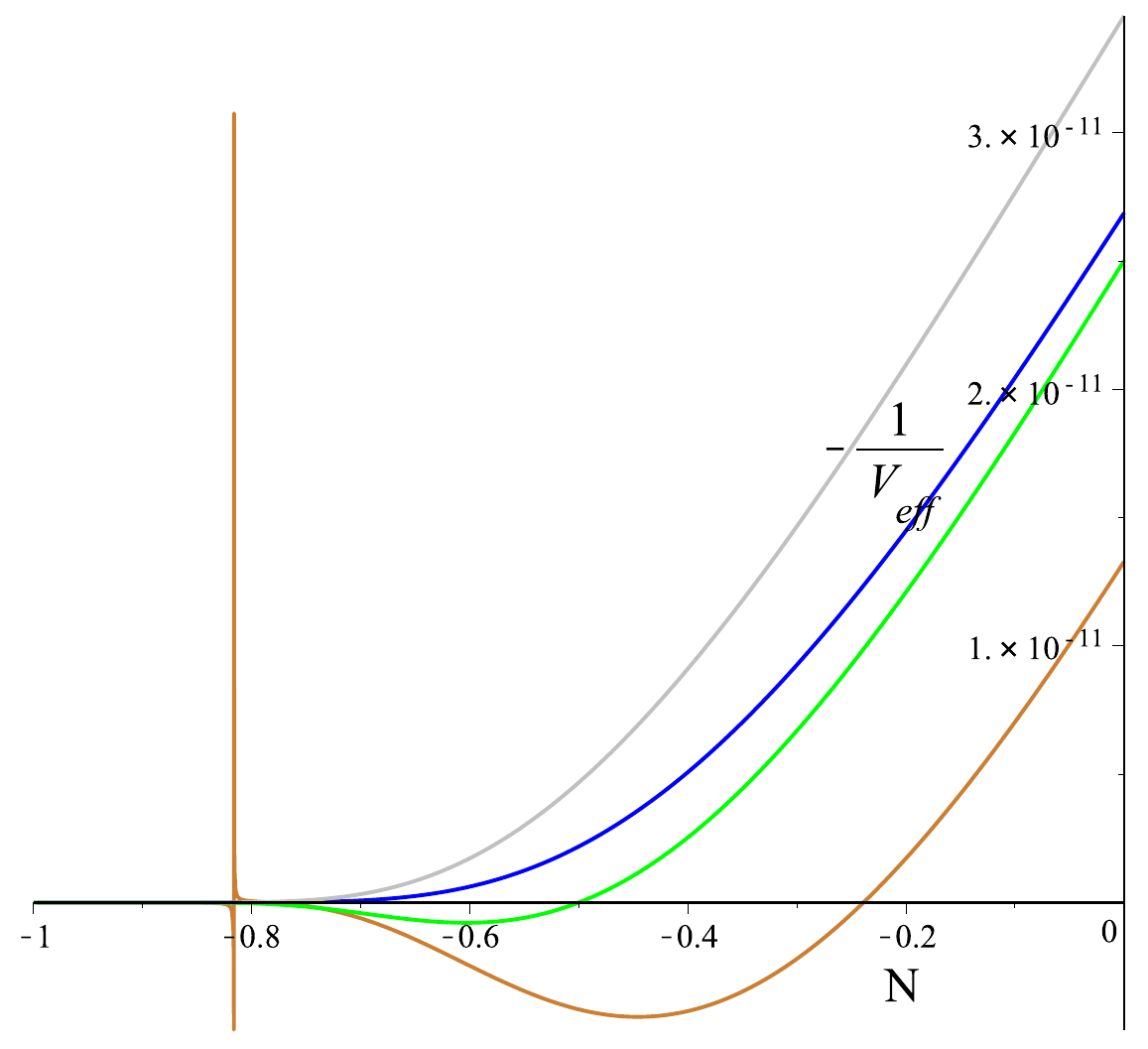}\quad
\caption{The behavior of $\tilde{V}_{eff}$  during inflation for slow-roll approximations (the gray line corresponds to the standard slow-roll approximation, the blue line -- to the approximation \eqref{V1}, the green line -- to the approximation \eqref{V2}) and the exact considerations (orange line) at the following values of the parameters: $N_0=1$, $N_b=57.787$, $\xi_0=1.6788\cdot10^{10}/\pi^2$, $Q_0\approx1.8861\cdot10^{-12}\pi^2$, $U_0={M_{Pl}^2}/{2}$, $M_{Pl}=1$. } \label{Veff(N)}
\end{figure}

All approximations and exact behavior of $\tilde{V}_{eff}$ have good agreements during inflation and start deviate near $N=5$. This deviation is mostly interesting after inflation, because after inflation only equivalent exact effective potential has potential well. The all approximations and exact expression of $\tilde{V}_{eff}$ are closed to zero after $N\approx-0.8$. The corresponding behaviors of the equivalent effective potential are presented in Fig.\ref{Veff(N)}.
The exact $\tilde{V}_{eff}$ has minimum at $N_m\approx-0.4463$ and equals to  $\tilde{V}_{eff}(N_m)\approx-4.4571\cdot10^{-12}$ (see the Fig.\ref{Veff(N)}). The equivalent effective potential $\tilde{V}_{eff}\approx0$ if $N\approx-0.2392$ and $N\approx-0.7777$.

Using the values of the model parameters, we obtain:
\begin{eqnarray}
  \Phi &=& (\phi^\prime)^2\approx\,{\frac {1.75 \left( N+ 1.5469 \right)  \left( N+ 0.73880 \right)
}{ \left( N+ 1 \right) ^{4}}},\\
  \phi &=&0.5\arctan\left(\frac{0.37796(N-1.7202\cdot10^{-9})}{S_q}\right)\\&+&1.3229\ln((N+1.1429+S_q)\cdot10^9)-\frac{1.3229\,S_q}{(N+1)}-c_0,\label{phiNUm}
\end{eqnarray}
where $S_q=\sqrt{(N+1.5468)\cdot(N+0.73886)}$, $c_0$ is a constant of integration. The choice of integration constant $c_0$ is not unique.
Near $N\approx-0.738$ the field becomes complex and we choose $c_0\approx25.431$ to get $\phi|_{N\approx-0.738}=0$.

\begin{figure}[htp]
\includegraphics[scale=0.25]{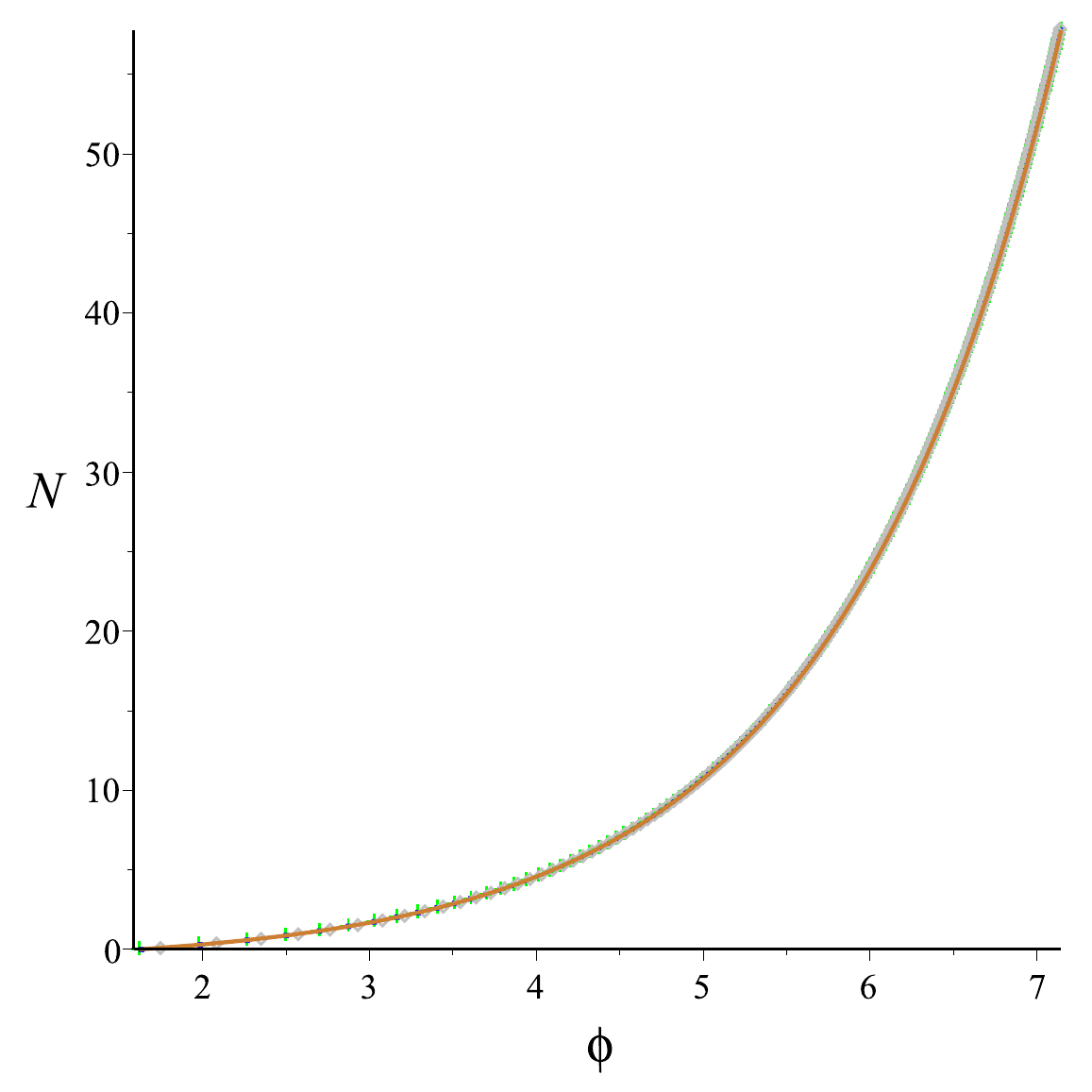}\quad
\includegraphics[scale=0.25]{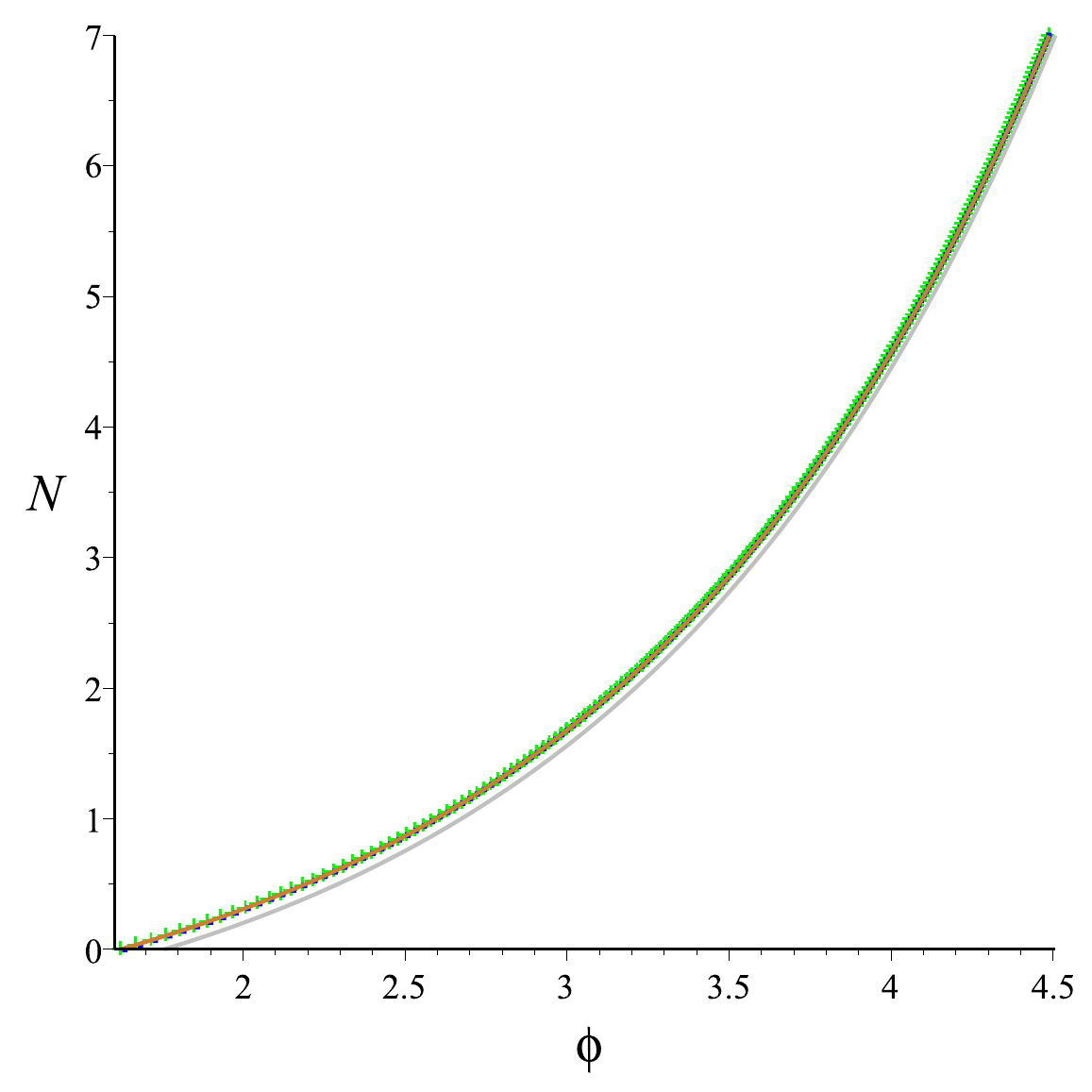}\quad
\includegraphics[scale=0.25]{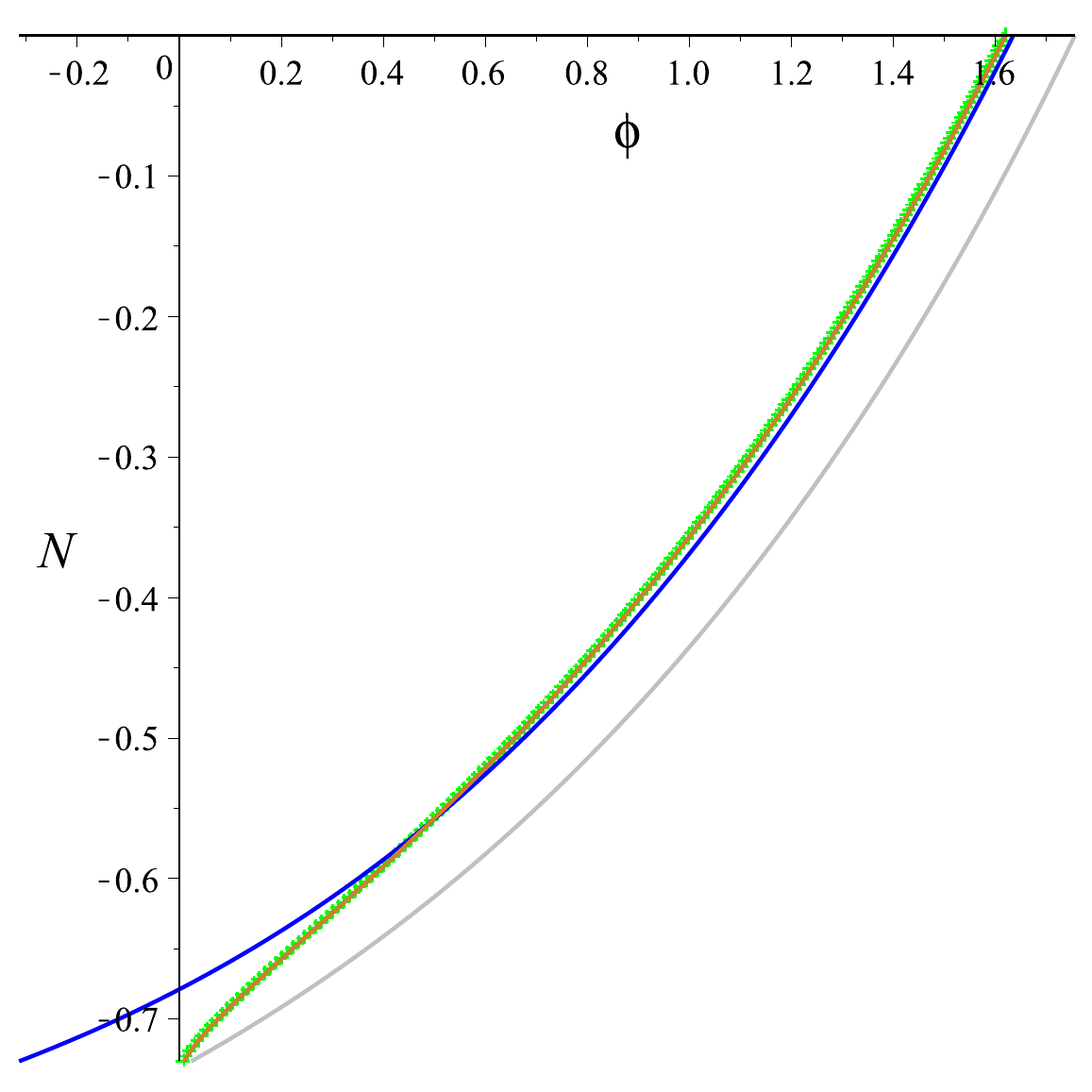}
\caption{The dependence of e-folding number form fields $N(\phi)$. The gray line corresponds to the standard slow-roll approximation, the blue line -- to the approximation \eqref{V1}, the green line -- to the approximation \eqref{V2} and the exact behavior at the following values of the parameters: $c_{sl}=1.7556$, $N_0=1$, $N_b=57.787$, $\xi_0=1.6569\cdot10^{10}/\pi^2$, $Q_0\approx1.8861\cdot10^{-12}\pi^2$, $U_0={M_{Pl}^2}/{2}$, $M_{Pl}=1$. In the left picture, all lines merge into one. In the central picture, the lines of extended approximations and exact behavior merge into one line again, the gray line has a small deviation from another lines near 5 e-folds before end of inflation. In the right, picture the first extended approximation deviates from the second extended approximation and the exact behavior after end of inflation} \label{N(phi)}
\end{figure}

We calculate the integration constant included to the slow-roll approximation of the field  $c_{sl}=\phi-(2\,{N_0}\,\sqrt {U_0-2\,{Q_0}\,{\xi_0}}\ln\left( N+N_0\right))$ at the point $N=N_b$ and get: $c_{sl}\approx1.75$. The choice of the integration constant allows us fix the same values of fields at the beginning of inflation.  We solve the differential equation $\frac{d\phi}{dN}=\sqrt{\Phi}$ numerically assuming $\phi(N=N_b)=7.145$ for all types of approximations and the exact solution. In Fig. \ref{N(phi)}, we can see that the graphical behavior of the field during inflation for the both slow-roll approximation and the exact solution has a small deviation near the $N=5$ e-folding number from the standard slow-roll approximation. The deviation increases after the end of inflation. The second extended slow-roll approximation coincides with exact behavior.

\begin{figure}[htp]
\includegraphics[scale=0.25]{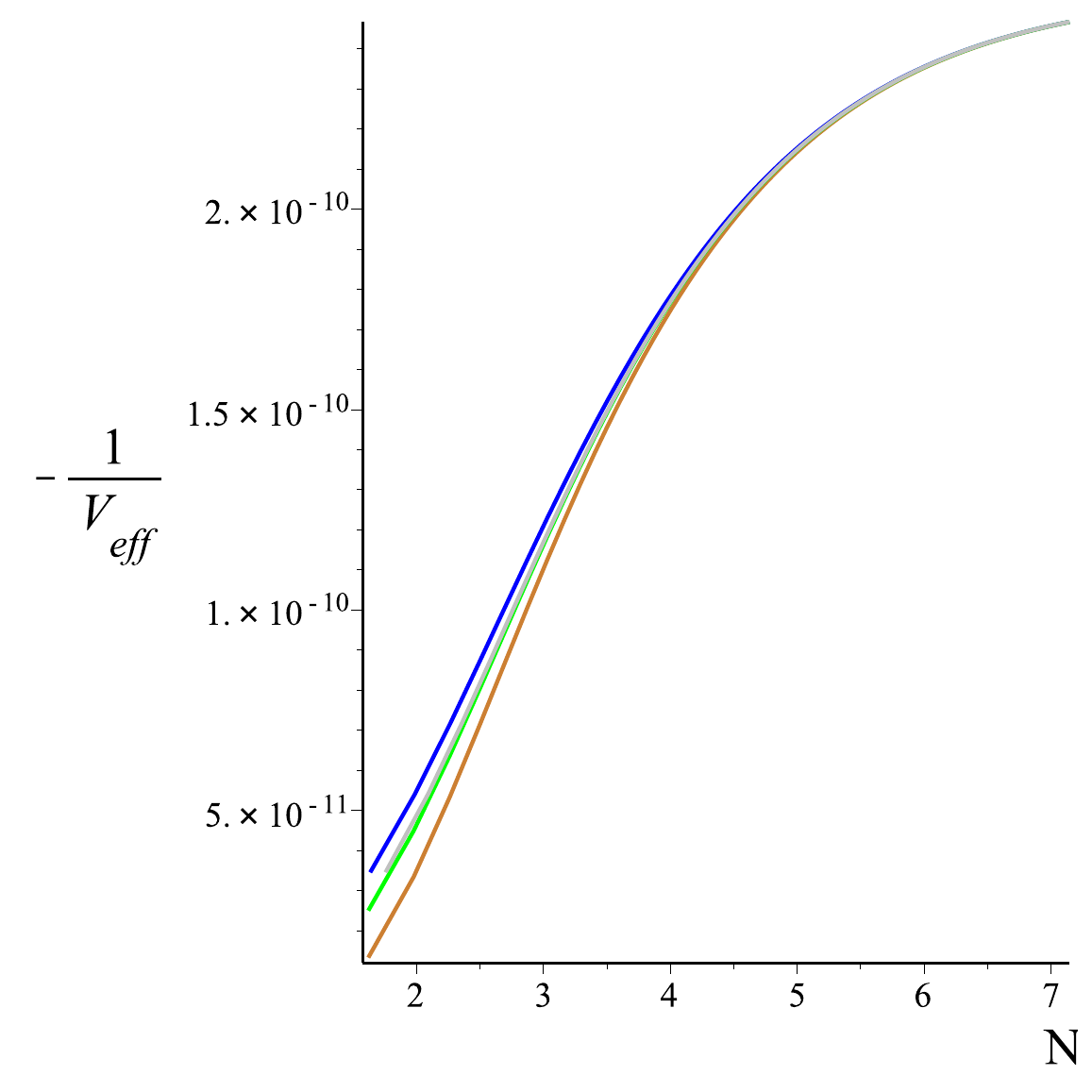}\quad
\includegraphics[scale=0.25]{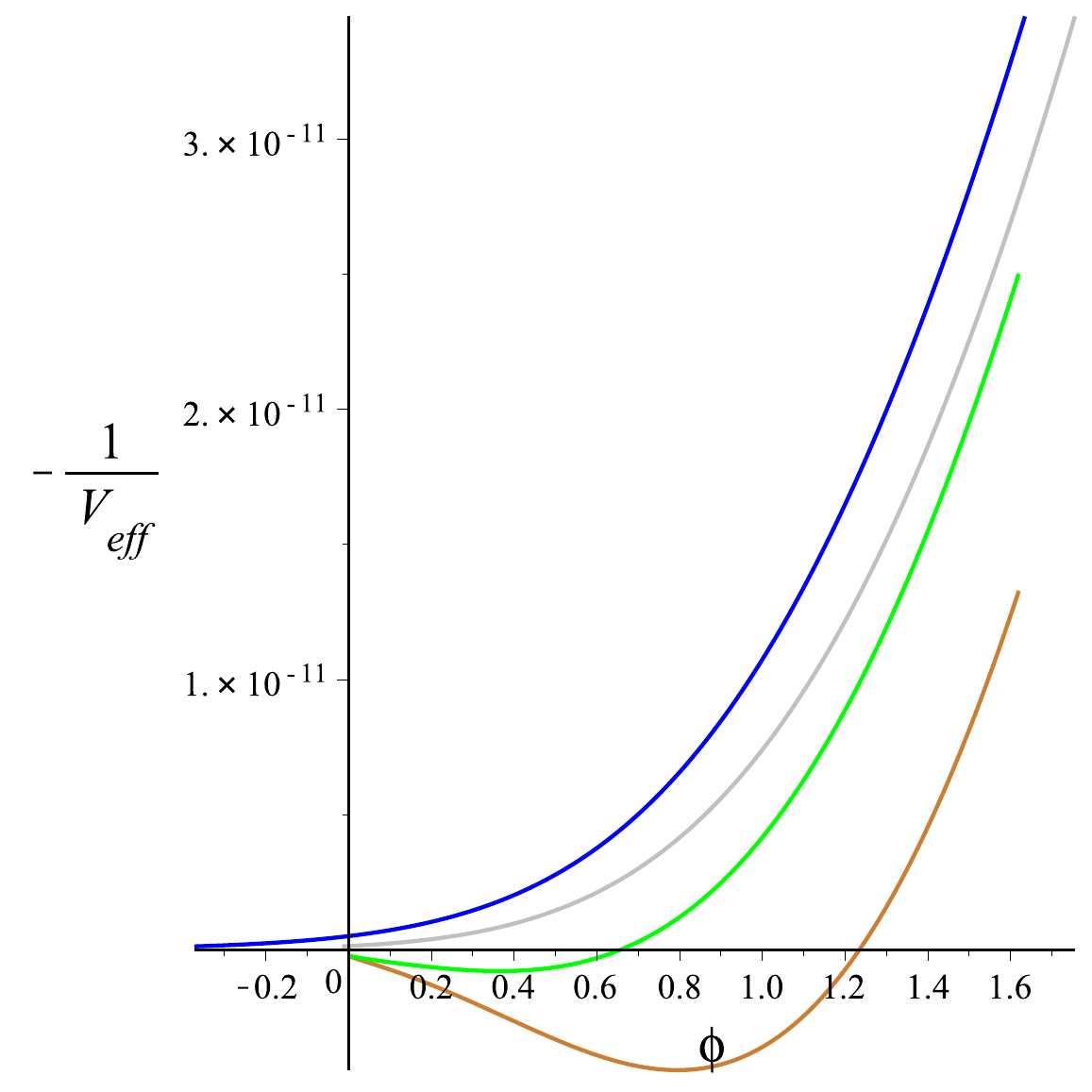}\quad
\includegraphics[scale=0.25]{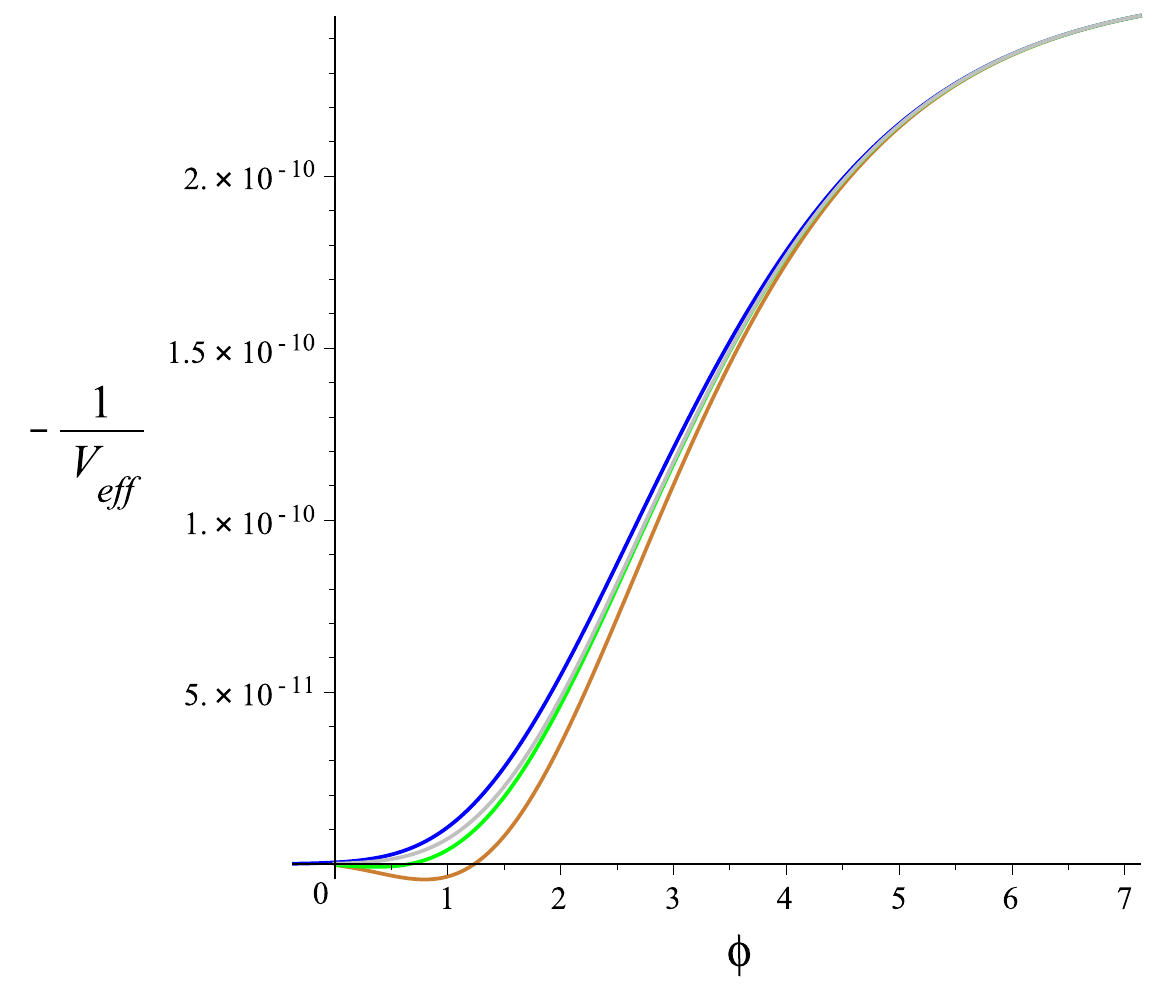}\quad
\caption{The graphical behavior of effective potential $\tilde{V}_{eff}=-V_{eff}^{-1}$  during inflation for slow-roll approximations (the gray line corresponds to the standard slow-roll approximation, the blue line -- to the approximation \eqref{V1}, the green line -- to the approximation \eqref{V2} ) and the exact considerations (orange line) at the following values of the parameters: $c_{sl}=1.7556$, $N_0=1$, $N_b=57.787$, $\xi_0=1.6569\cdot10^{10}/\pi^2$, $Q_0\approx1.8861\cdot10^{-12}\pi^2$, $U_0={M_{Pl}^2}/{2}$, $M_{Pl}=1$. The left picture corresponds to evolution during inflation, the middle picture - to evolution after inflation, the right picture is the join evolution.} \label{Veff(phi)}
\end{figure}

In Fig. \ref{Veff(phi)}, the dependence of the equivalent effective potential $\tilde{V}_{eff}$ on the field for all type of approximation and exact case are presented.

 The exact dependencies of equivalent effective potential from e-folding number $\tilde{V}_{eff}(N)$ and field $\tilde{V}_{eff}(\phi)$ have potential well after inflation, but approximations of equivalent effective potential don't have such potential well. At the same time all approximations rather accurate reproduce the exact behavior up to $N\approx5$. So, we suppose that for the considering model the application of slow-roll approximation and it's extended version is reasonable for description of inflation. And it is reasonable to apply the exact behavior simulation to better understanding processes after inflation.

In Appendix \ref{Pictures} we demonstrate how different types of approximations influence to value of field corresponding to end of inflation. We present graphics of  $Q\cdot\xi$, $V\cdot\xi$ from field $\phi$ and e-folding numbers. The behavior of Gauss-Bonnet interaction with scalar field $\xi(\phi)$ is presented in Appendix \ref{Pictures} too.
\subsection{Wave speeds}\label{Wave speeds}
\begin{figure}[htp]
\includegraphics[scale=0.2]{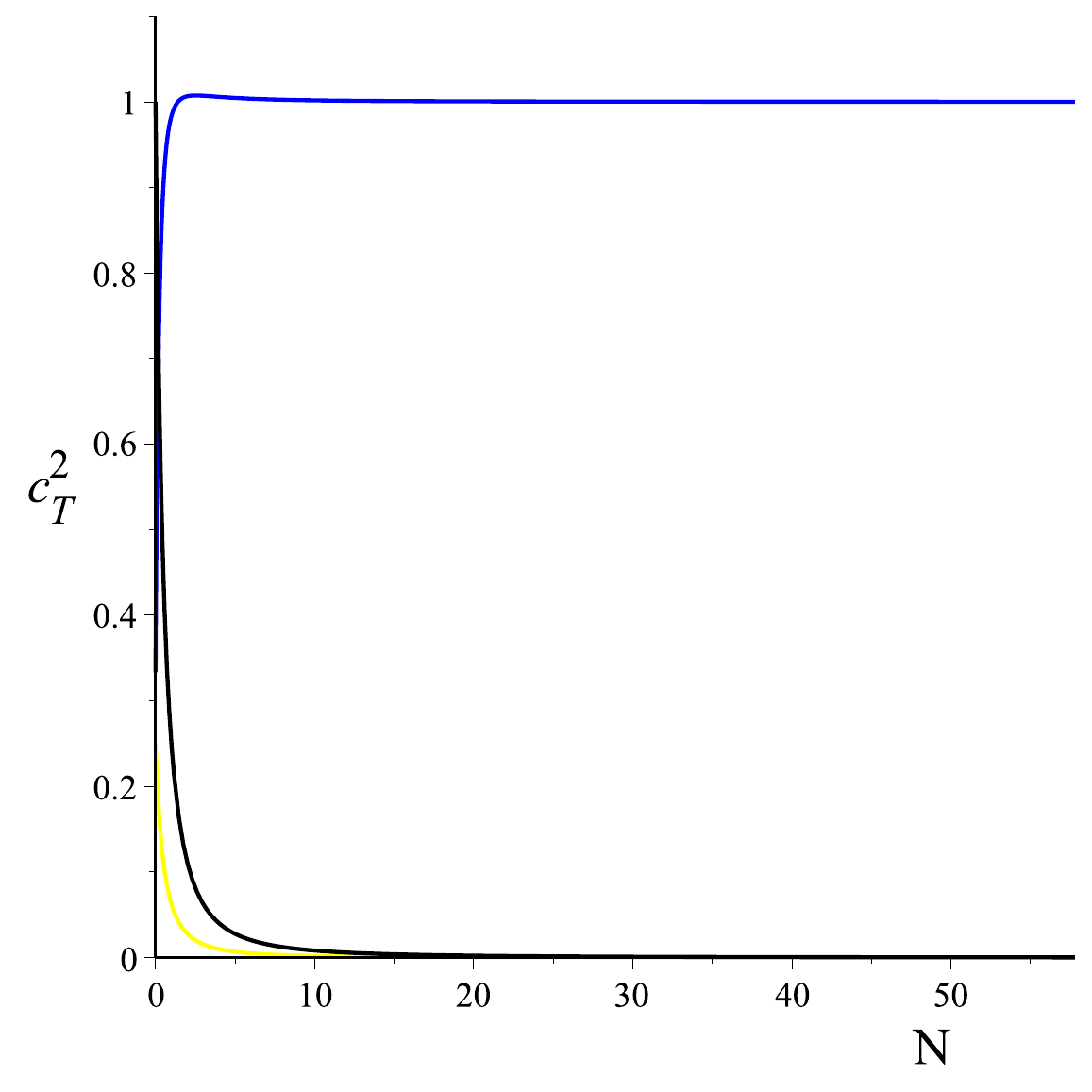}\,
\includegraphics[scale=0.2]{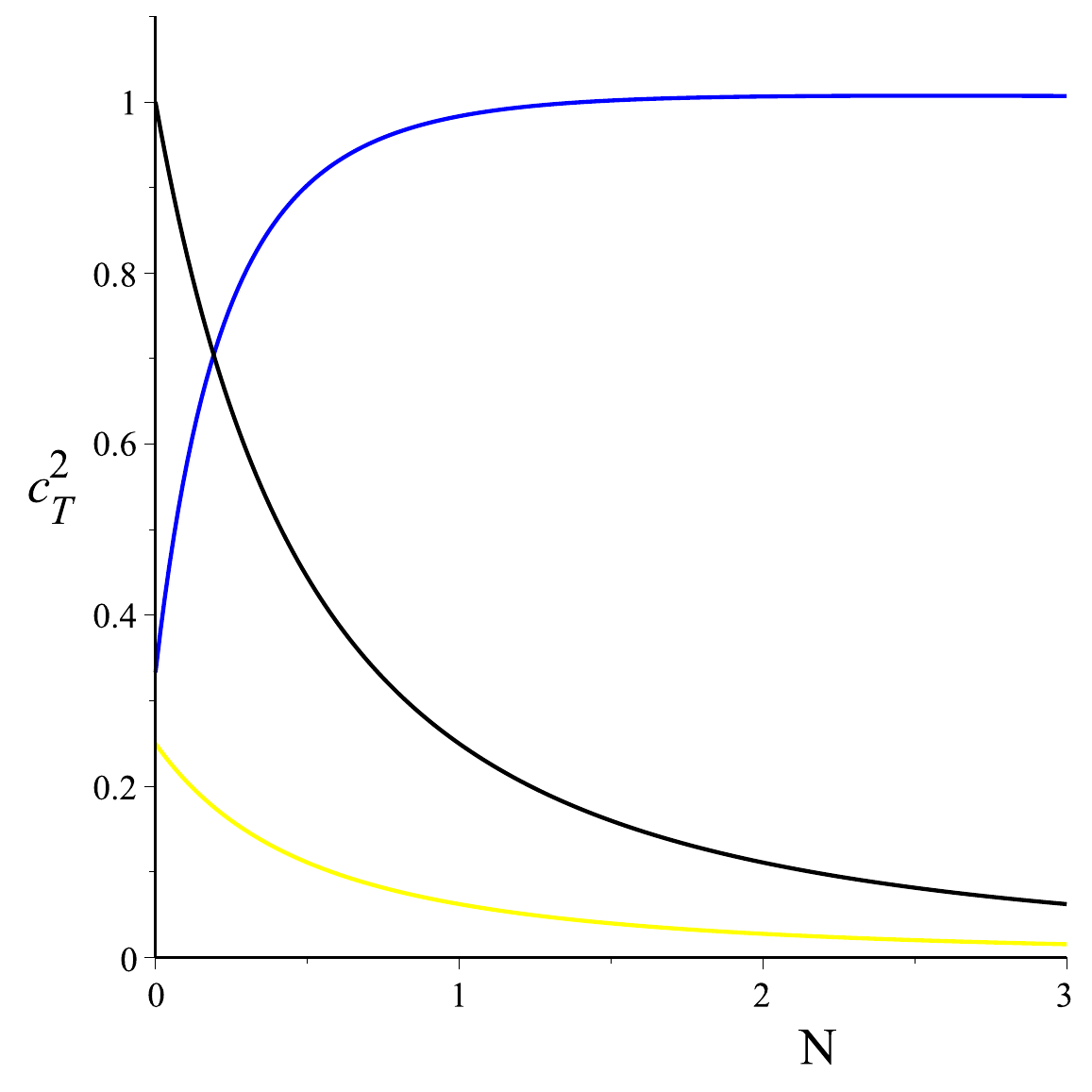}\,
\includegraphics[scale=0.2]{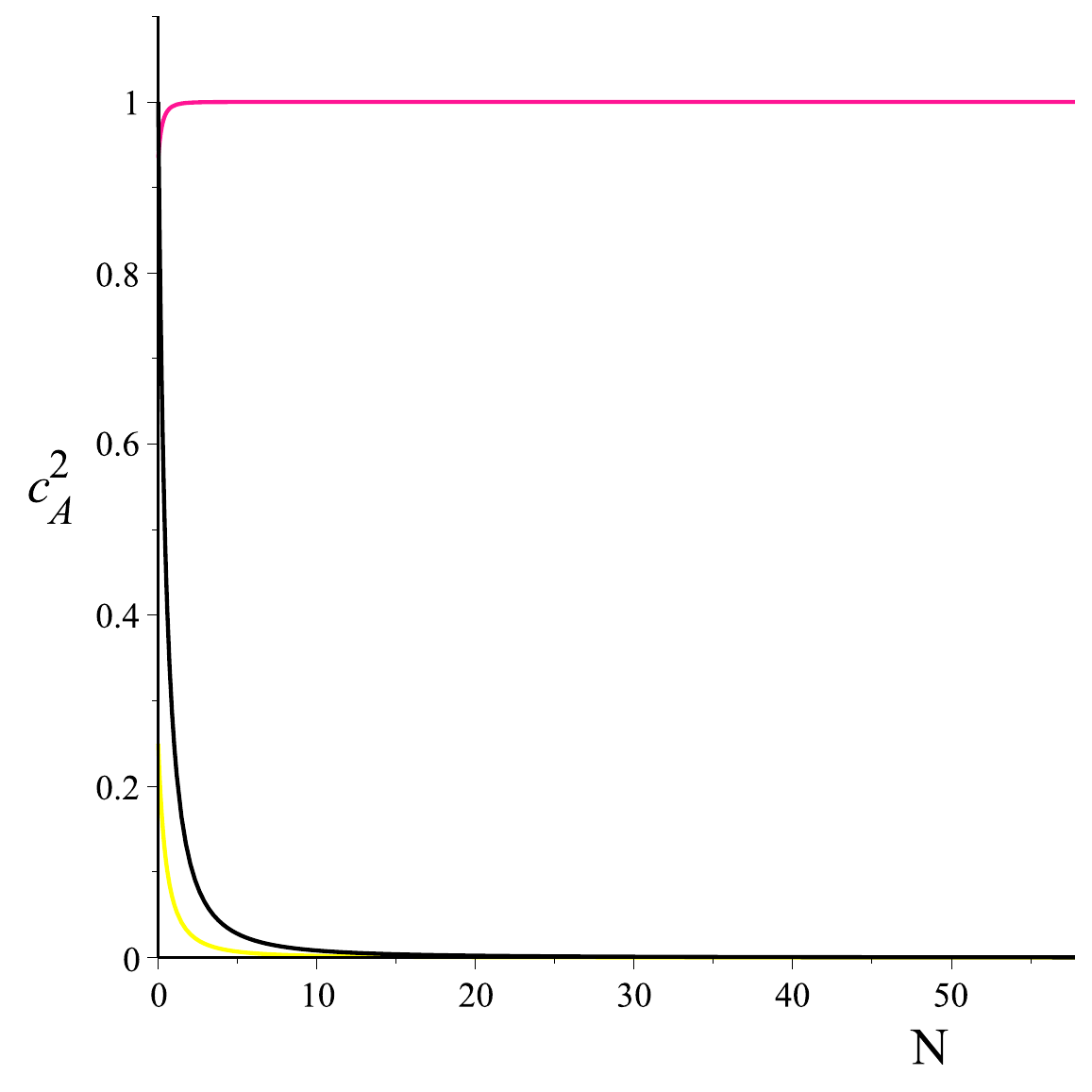}\,
\includegraphics[scale=0.2]{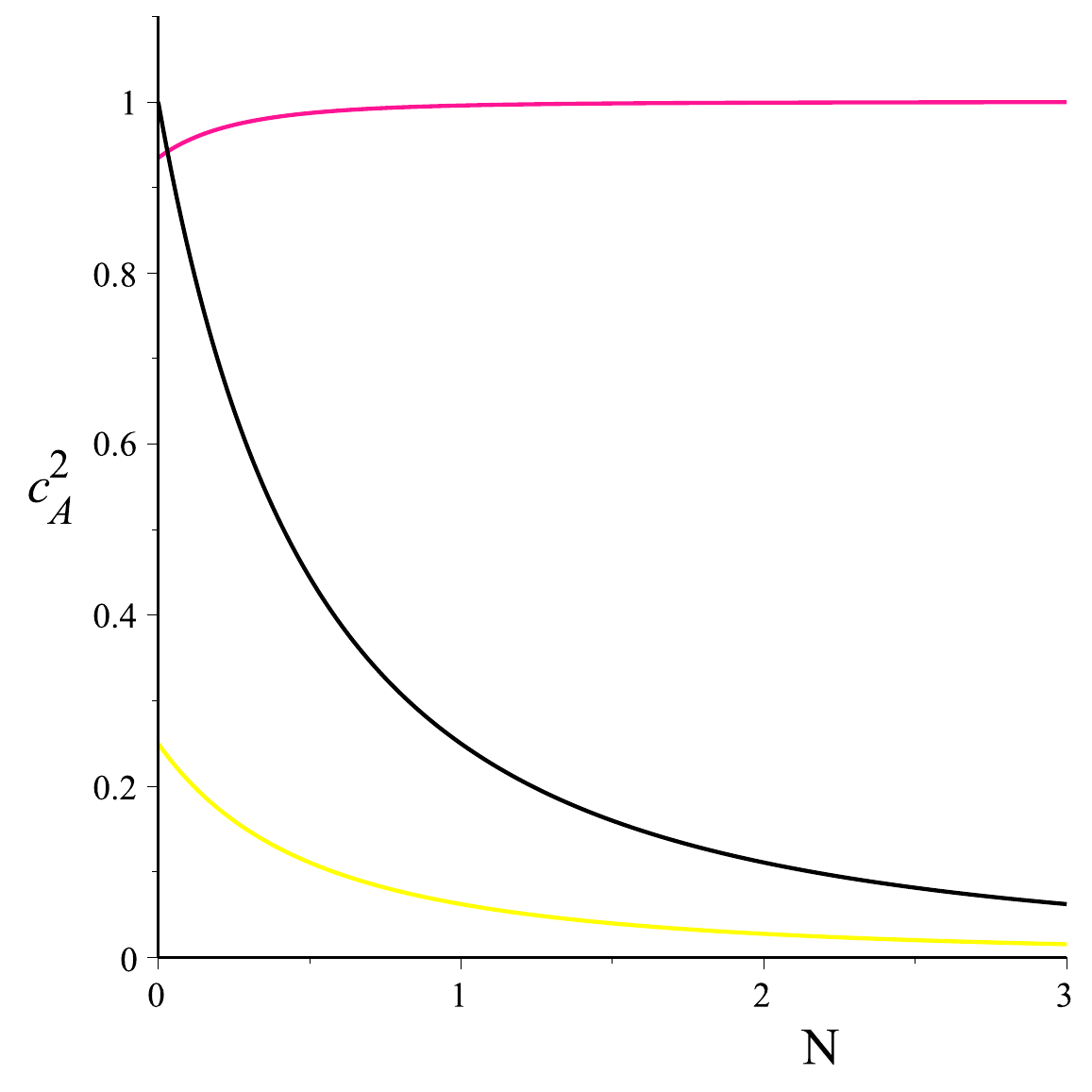}
\caption{There are slow-roll parameters $\epsilon_1$ (black lines), $\delta_1$ (yellow lines)  and  squares of the wave speed of the perturbed field $c^2_A$ (magenta line), speed of GW (blue line)  at the following values of parameters: $N_0=1$, $N_b=57$, $\xi_0\approx1.6192\cdot10^{10}/\pi^2$, $Q_0\approx1.9299\cdot10^{-12}\pi^2$, $U_0={M_{Pl}^2}/{2}$, $M_{Pl}=1$.} \label{C2}
\end{figure}

We calculate squares of the wave speed of the perturbed field and
the speed of gravitational wave using formulas \cite{Hwang:2005hb} in terms of derivatives
with respect to e-folding numbers:
\begin{eqnarray}
c^2_A&=&1+\frac{Q_a\,Q_e/(4\,U_0+Q_b)+Q_f\,Q_a^2/(4\,U_0+Q_b)^2}{Q\,\Phi+3\,Q_a^2/(4\,U_0+Q_b)}; \\
c^2_T&=&1-\frac{Q_f}{4\,U_0+Q_b}
\end{eqnarray}
where $Q_a=4\,Q^{3/2}\,\frac{d\xi}{dN},$ $Q_b=8\,Q\,\frac{d\xi}{dN},$
$Q_e=-16\,Q\,\frac{d\xi}{dN}\frac{d\,Q^{1/2}}{dN},$ $Q_f=
8\left(\left(\frac{1}{2}\,\frac{d Q}{dN}+Q\right)\frac{d\xi}{dN}+Q\,\frac{d^2\xi}{dN^2}\right).$
In Fig. \ref{C2} graphical behavior of $c^2_T$ and $c^2_A$ are presented using expressions and constants from the text above.
The $c^2_A$, $c^2_T$ limit $1$ with growing of $|N|,$ so at present $c^2_A\approx 1$, $c^2_T\approx 1$. The sufficient deviation of $c^2_A$, $c^2_T$ from unity coincides with increasing of parameter $\delta_1$. It can mean the inapplicability of the early obtained formulas due to increasing of perturbations at some values of e-folding numbers. Moreover, using the dependence of exact equivalent effective potential we can conclude that the considering model does not work after $N\approx-0.8$
and during investigation of later processes the General Relativity model should be applied. So, because the speed of GW was measured much later than inflation ended,
we can suppose that the considering model does not contradict any GW170817 event results.
\subsection{Spectral index of tensor perturbations}\label{Specter of tensor perturbations}
In the text above the value of inflationary parameters are restricted due to modern observations \cite{Planck:2018jri,BICEP:2021xfz,Galloni:2022mok}. The other inflationary parameters can exist too, for example, the spectral index of tensor-type perturbations \cite{Hwang:2005hb}. We estimate value of $n_T$  using expression
\begin{equation}
n_T\approx\frac{d\ln(Q(2\,U_0+Q_b/2))}{dN}={\frac {2\,{{N_0}}^{2}}{ \left( N+{N_0} \right)^{2}}}+{
\frac {8\,{\xi_0}\,{{N_0}}^{2}{Q_0}}{ \left( N+{N_0} \right)
 \left(\left( {(N+N_0)}^{2}\right) {U_0}-4\,{\xi_0}\,{{N_0}}^{2}{Q_0}\right) }}
\end{equation}
in the beginning of inflation $N=N_b$ and get:
\begin{equation}
n_T\approx0.00058.
\end{equation}
The value of the spectral index of tensor perturbations is positive and increases during inflation to $n_T\approx2.7$ at the end of inflation. Accordingly with \cite{Stewart:2007fu,Oikonomou:2024aww} the obtained $n_T$ the spectrum of gravitational waves is tilted to the blue. It is interesting to note, the blue tilted spectrum of tensor perturbation can lead to production of particles \cite{Wang:2014kqa}. The behavior of $n_T(N)$ during inflation is presented in Fig. \ref{C2}. The Hubble parameter decreases during inflation (see Fig.\ref{C2})
\begin{figure}[htp]
\includegraphics[scale=0.25]{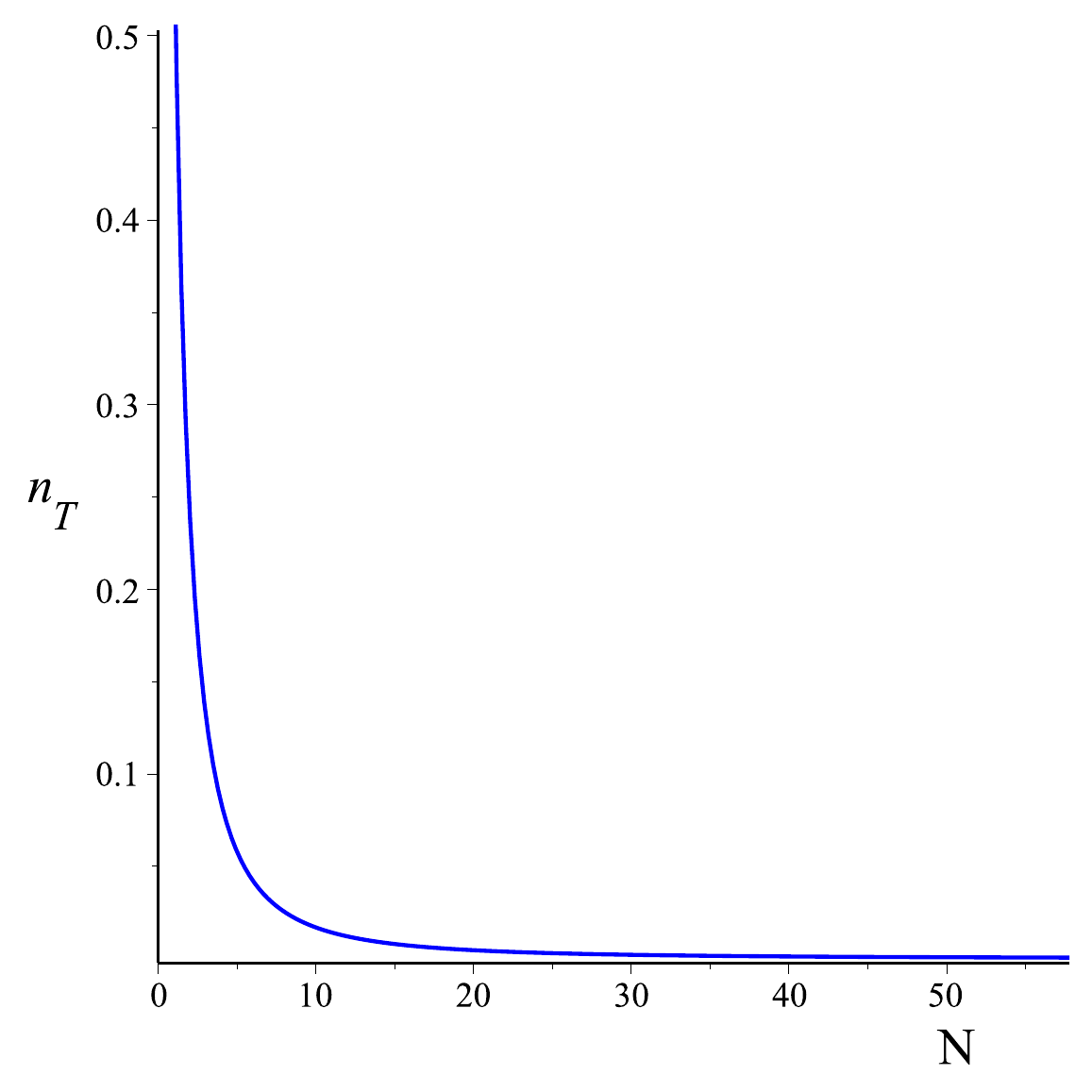}
\includegraphics[scale=0.25]{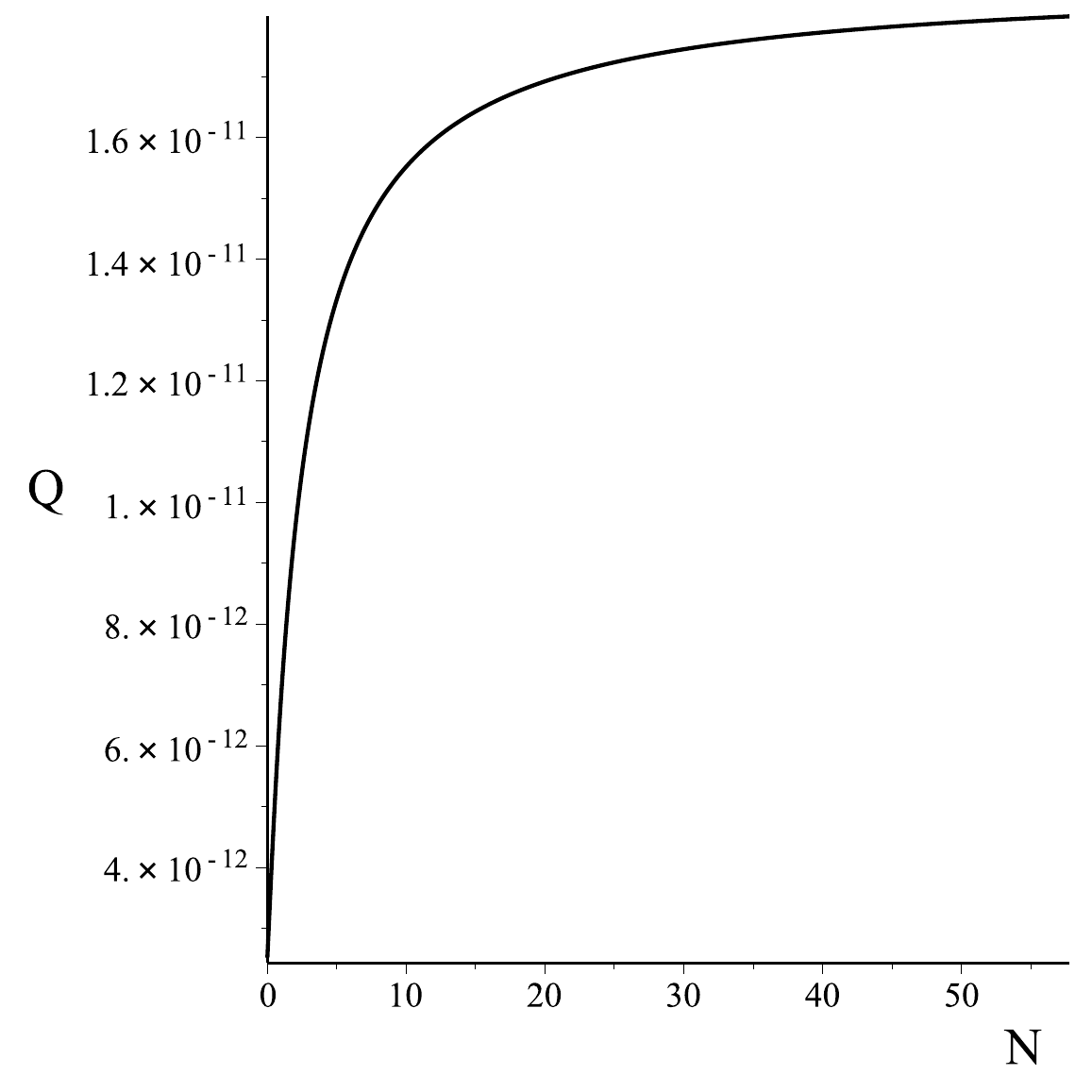}\,
\includegraphics[scale=0.25]{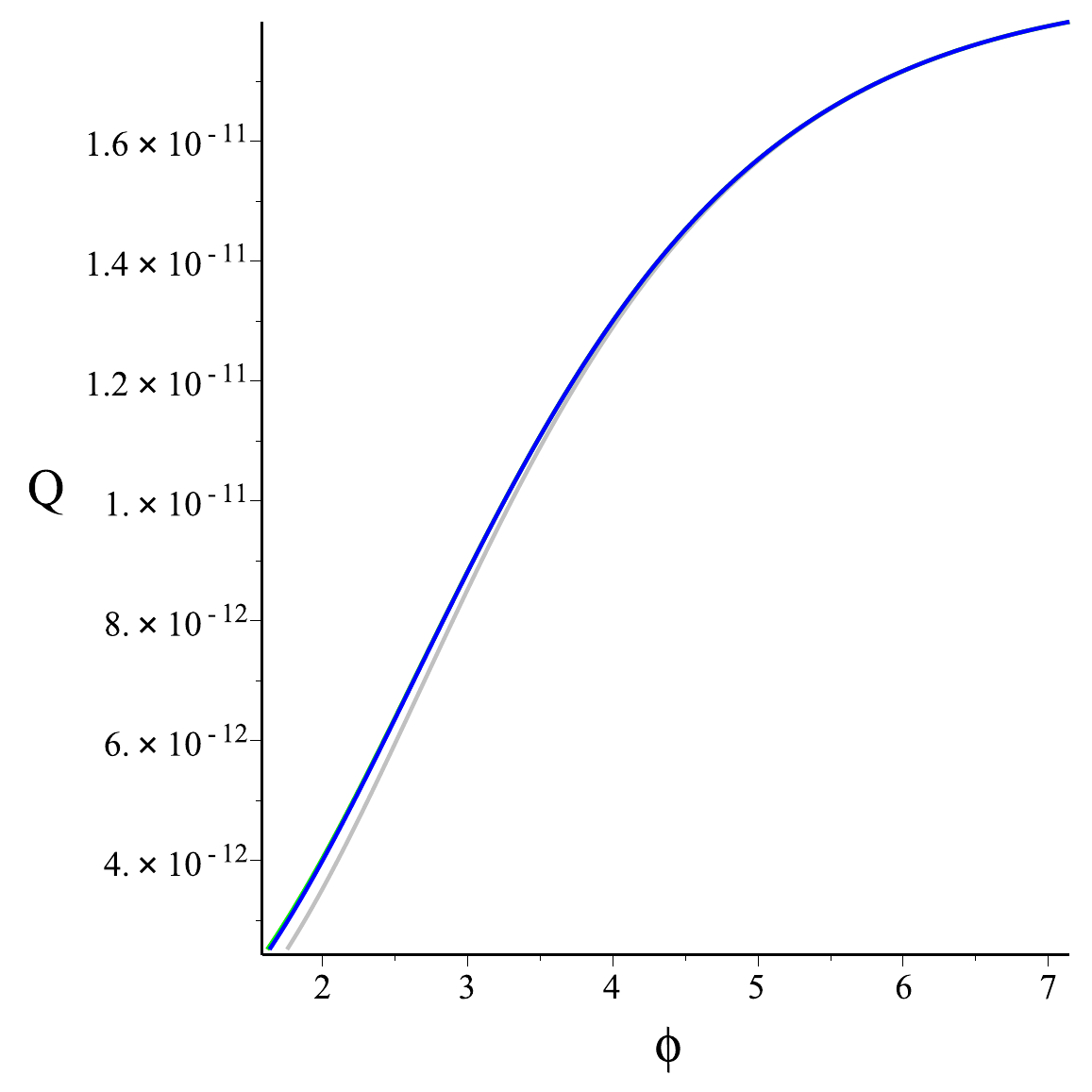}\,
\caption{The behavior $n_T(N)$ during inflation is in left picture, the graphical dependence of $H^2(N)$ is in the central picture, $H^2(\phi)$ is in the right picture at the following values of parameters: $N_0=1$, $N_b=57$, $\xi_0\approx1.6192\cdot10^{10}/\pi^2$, $Q_0\approx1.9299\cdot10^{-12}\pi^2$, $U_0={M_{Pl}^2}/{2}$, $M_{Pl}=1$.} \label{nT}
\end{figure}
\section{Conclusion}\label{Conclusion}
The standard and the extended slow-roll approximations were considered and verified using the exact solution for the exponential model of EGB gravity.
The extended slow-roll approximation with $V\sim(1-\delta_1)$ allows us to reconstruct the dependence of exact field on the e-folding numbers. However,
the extended slow-roll approximation with $V\sim(1+\delta_1)^{-1}$ does not lead to an analytical dependence of the field on the e-folding numbers.
All potentials reconstructed in this study preserve the exponential nature of the potential obtained within the standard slow-roll approximation framework. The first terms of all potentials coincide with $V_{sl}$. The standard slow-roll approximation reproduces $\phi(N)$ of the exact model for $N$ between $N\approx58$ and $N\approx5$ up to an integration constant. After that, the deviation slowly increases before the end of inflation.
Therefore the substitution of the standard slow-roll approximated $\phi(N)$ instead of the exact consideration to the models depending from
field can change the total number of inflationary e-folding number in numerical tests. Despite this the  field $\phi(N)$ approximated by standard slow-roll can be used to generate new EGB inflation models due to its simplicity in reasonable values of e-folding number.
Both the standard and the extended slow-roll approximations reproduce exact effective potential up to $N\approx5$ e-folding number with hight accuracy.
The effective potential obtained from the standard slow-roll approximation has bigger deviation from exact than obtained from the extended slow-roll approximations. We suppose that in models with difficult exact analytical considerations, the extended approximation with $V\sim(1-\delta_1)$  can be applied with higher accuracy to descriptive inflation. The results obtained in this paper can be applied for generation of EGB gravity models. We hope that the results of the paper can be useful for better understanding of post-inflation evolution and for connection of early and later time processes \cite{TerenteDiaz:2023iqk,Pinto:2024dnm} to general models.\\

Graititude: `The study was conducted under the state assignment of Lomonosov Moscow
State University'
\appendix\section{Appendix}\label{Appendix}
\subsection{Approximations}\label{potential}
The equations \eqref{leads to V} can be presented in the form:
\begin{eqnarray}
  & & 6\,U_0\,H^2\,\left(1-\delta_1\right)=\frac{\dot{\phi}^2}{2}+V.\label{delta of leads to V}
\end{eqnarray}
Supposing $\delta_1\ll1$ and $\dot{\phi}^2/2\ll6U_0H^2$ one can get the standard slow-roll approximation:
\begin{equation}
V\approx6\,U_0\,Q.
\end{equation}
The slow-roll parameter $\delta_1$ grows during inflation and can becomes equal to one before end of inflation and can be taken into account
considering slow-roll approximation of \eqref{leads to V}:
\begin{equation}
6U_0\,Q(1-\delta_1)\approx\,V. \label{extended leads to V}
\end{equation}
From here we get the potential approximations \eqref{V2} and \eqref{V1} using relation
\begin{equation}
(1-\delta_1)\approx\frac{1}{1+\delta_1}.
\end{equation}
\subsection{Pictures}\label{Pictures}
\begin{figure}[htp]
\includegraphics[scale=0.2]{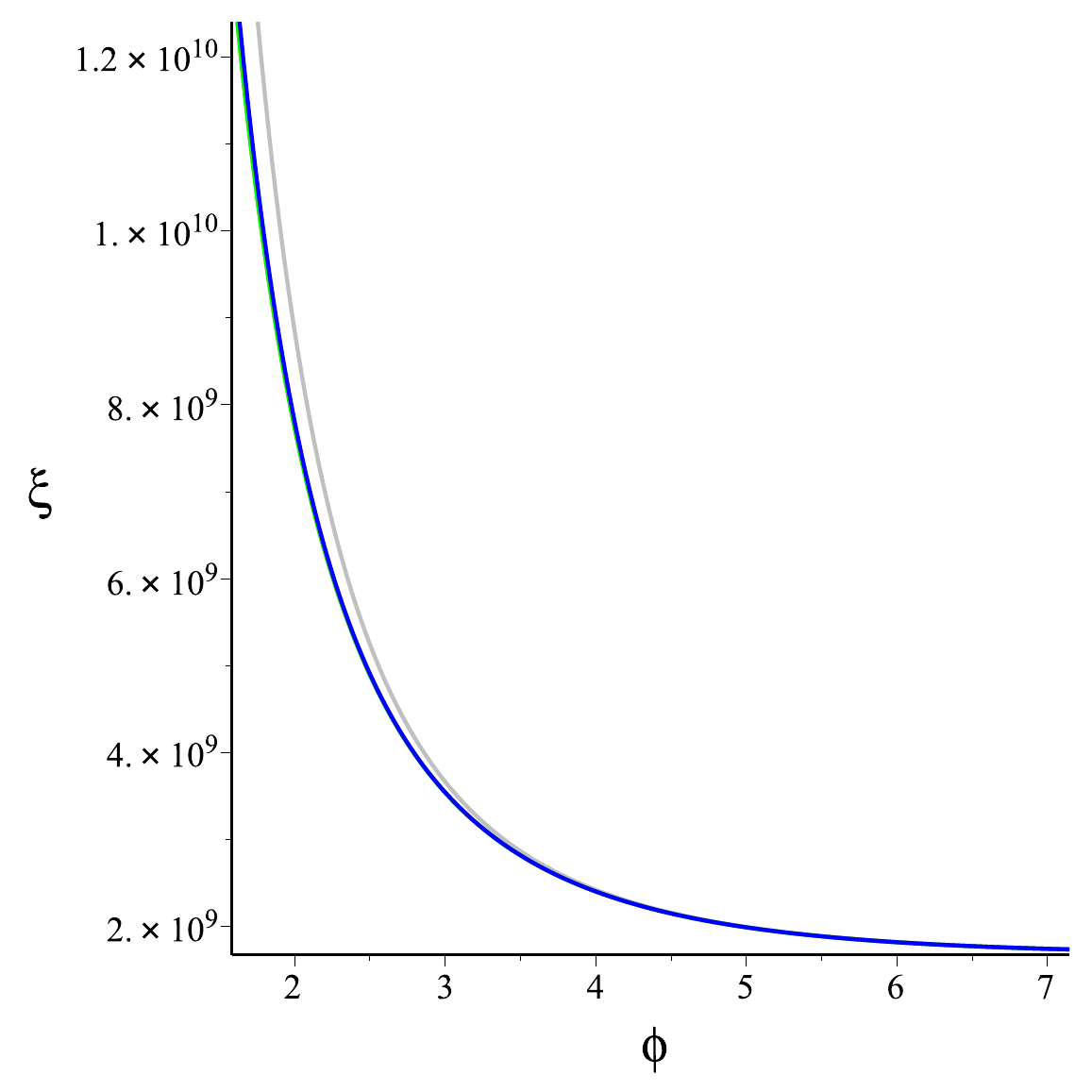}\,
\includegraphics[scale=0.2]{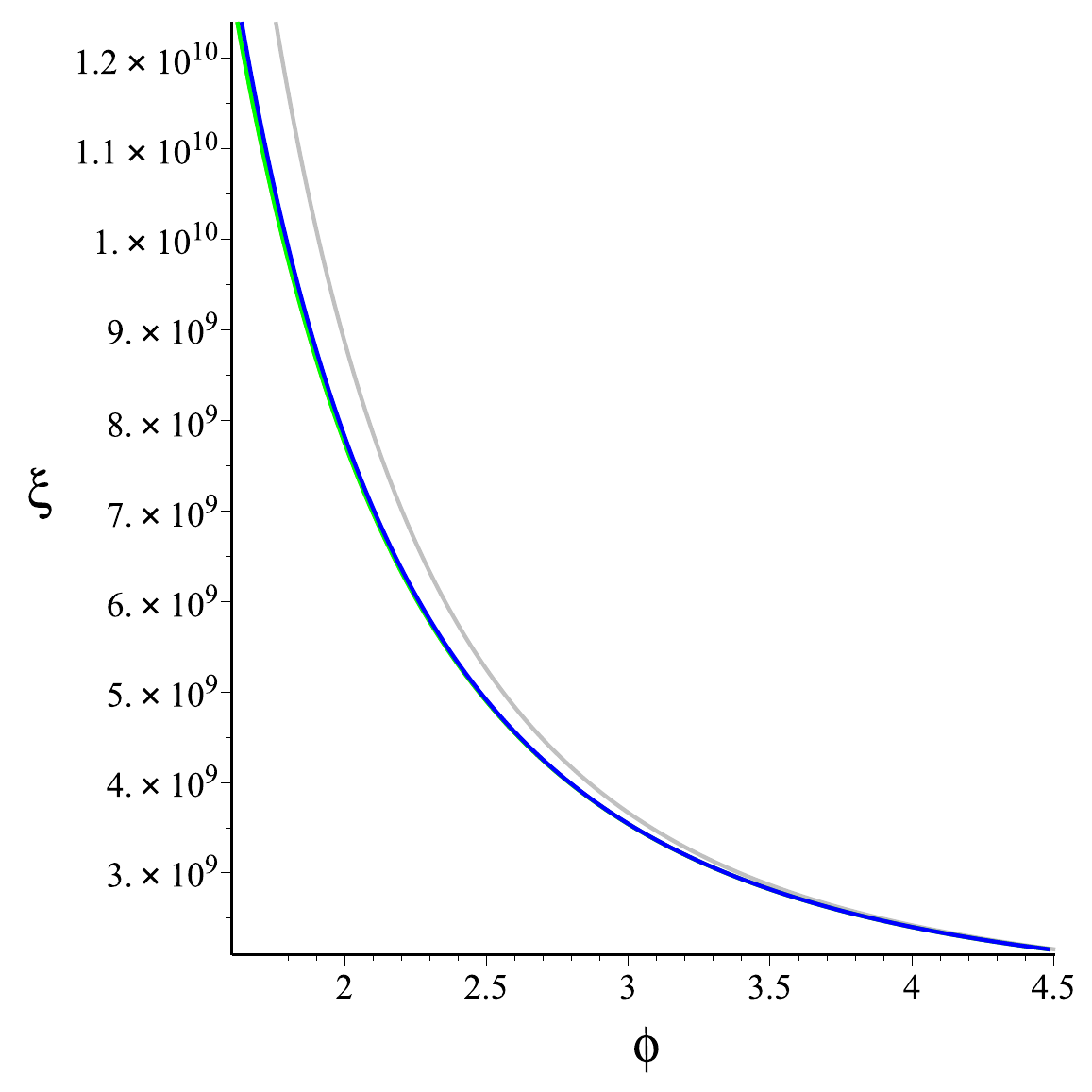}\,
\includegraphics[scale=0.2]{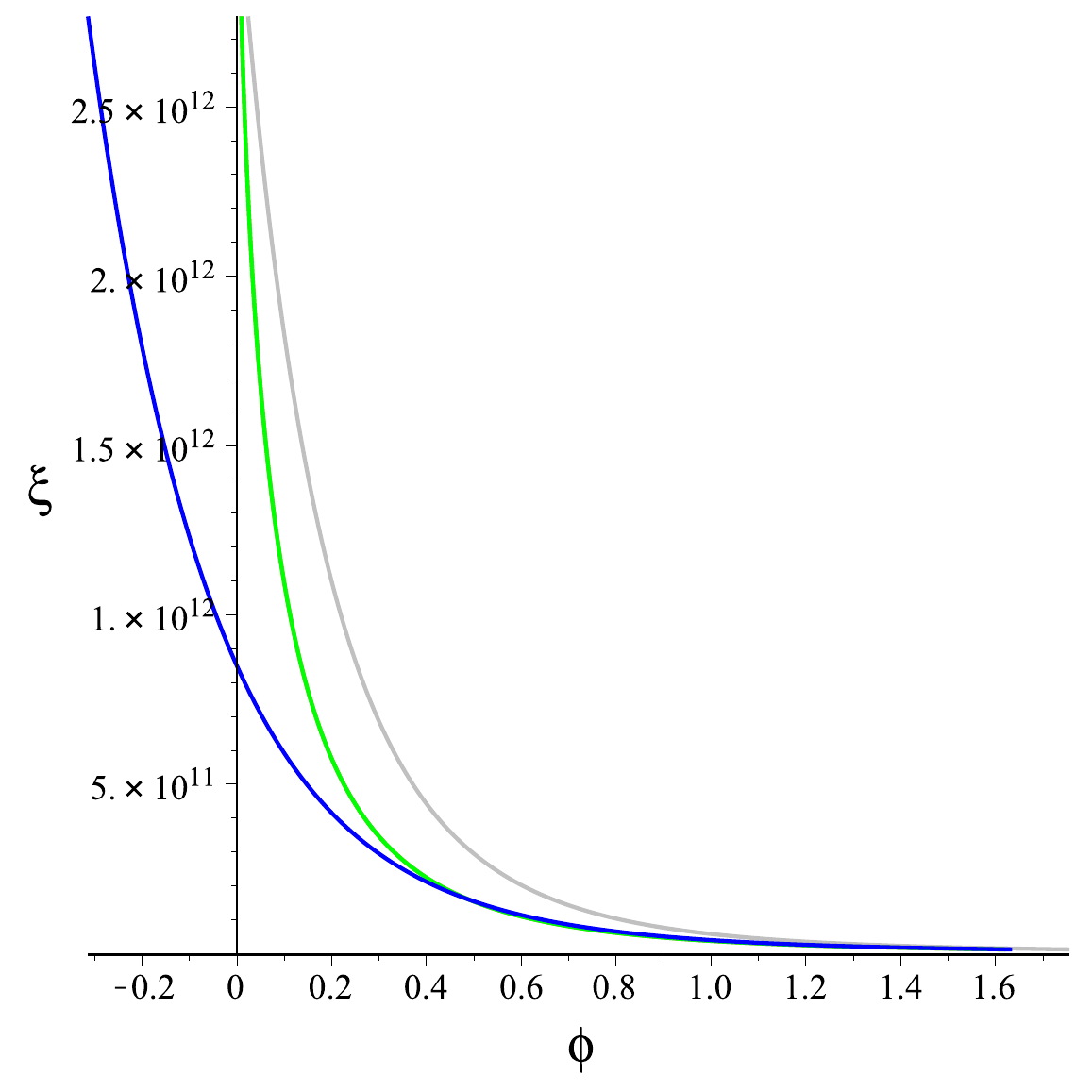}\,
\includegraphics[scale=0.2]{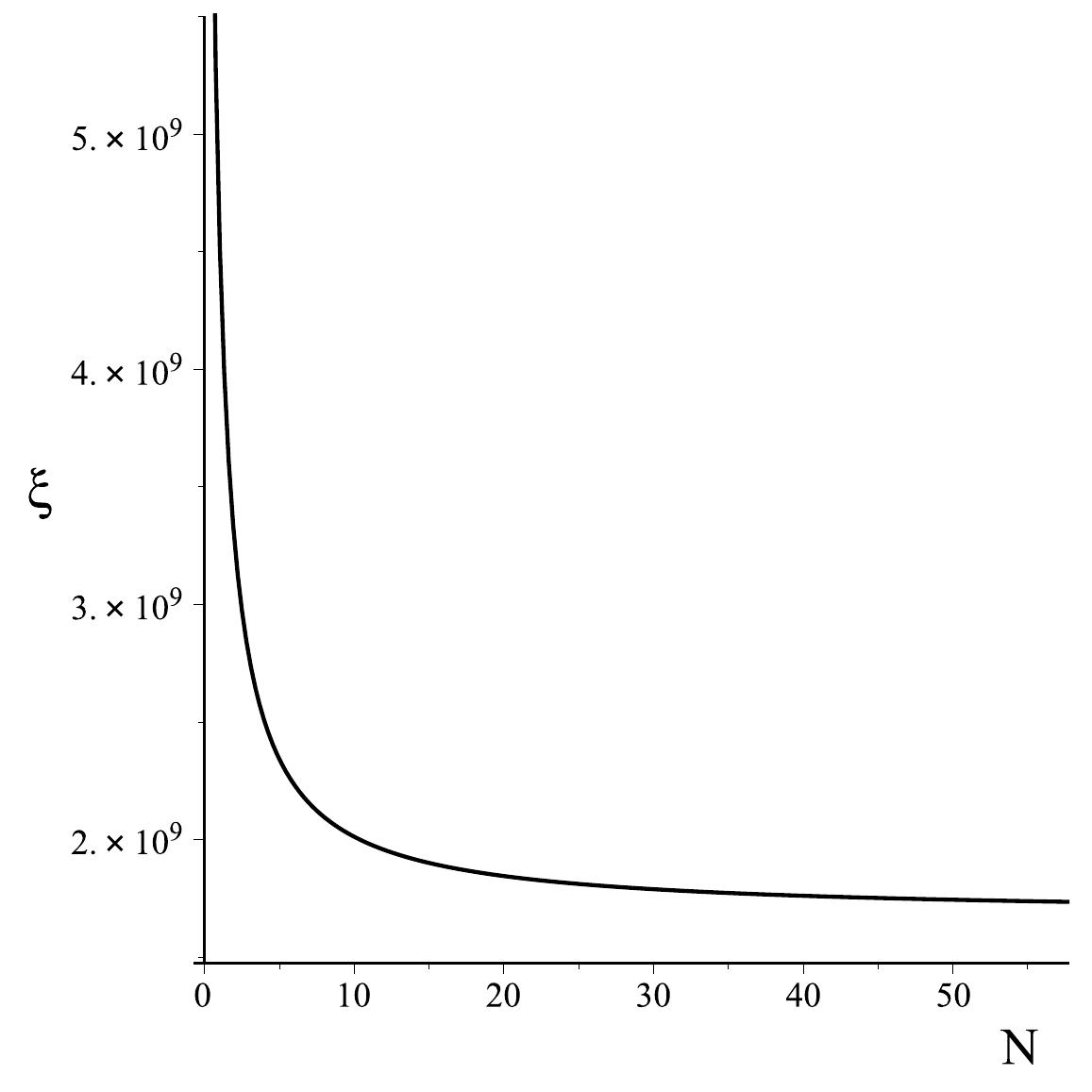}\,
\caption{The  Gauss-bonnet coupling function $\xi(\phi)$ is in first three pictures (the gray line corresponds to the standard slow-roll approximation, the blue line -- to the approximation \eqref{V1}, the green line -- to the approximation \eqref{V2},  the exact consideration coincides with green line)   from $N=0..N_b$, from $N=0..7$ and from $N=-0.73..0$. The graphical behavior of Gauss-Bonnet coupling $\xi(N)$  from $N=-0.73..N_b$ is in fourth picture. We use the following values of the parameters: $c_{sl}=1.7556$, $N_0=1$, $N_b=57.787$, $\xi_0=1.6569\cdot10^{10}/\pi^2$, $Q_0\approx1.8861\cdot10^{-12}\pi^2$, $U_0={M_{Pl}^2}/{2}$, $M_{Pl}=1$.} \label{xi(phi)}
\end{figure}

\begin{figure}[htp]
\includegraphics[scale=0.25]{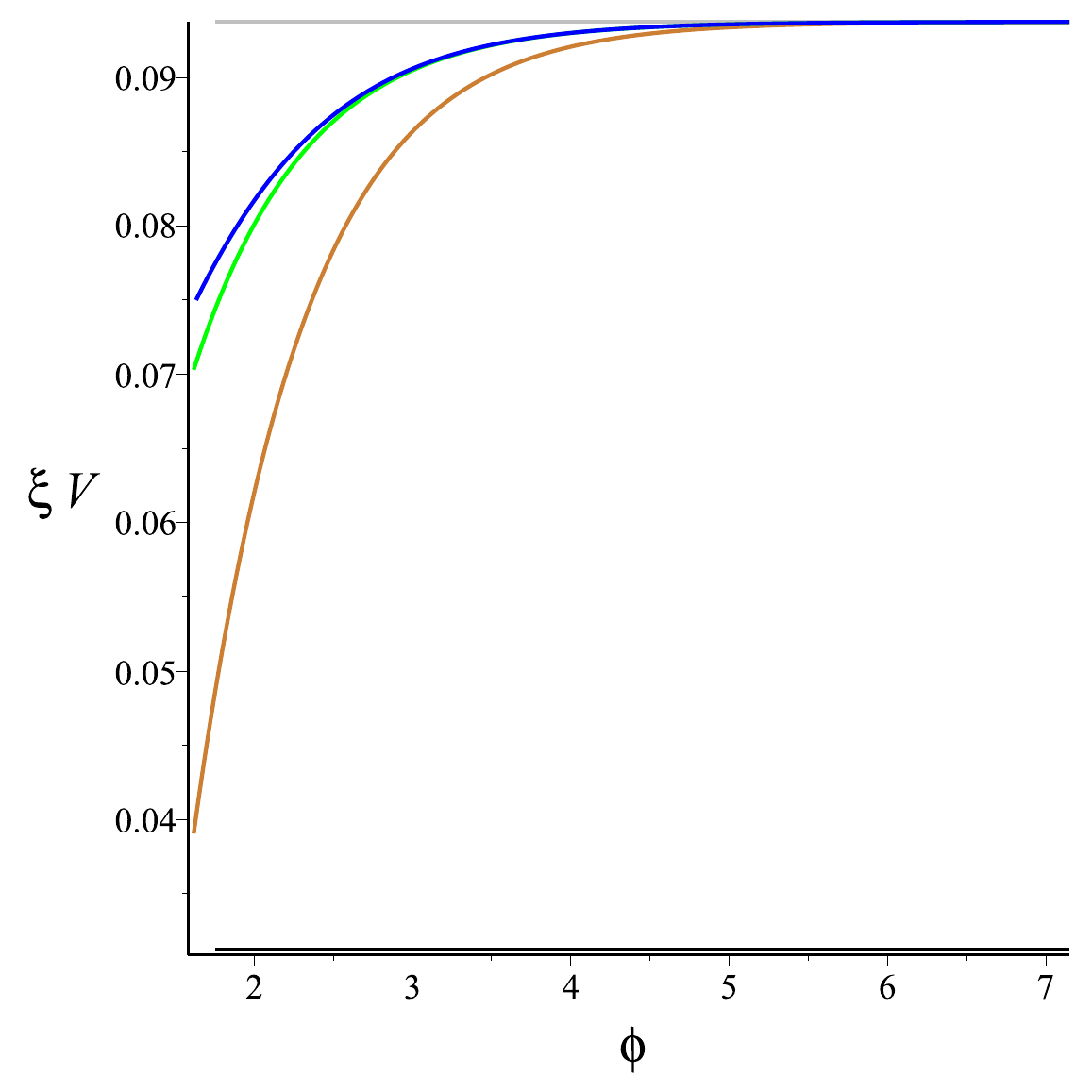}\,
\includegraphics[scale=0.25]{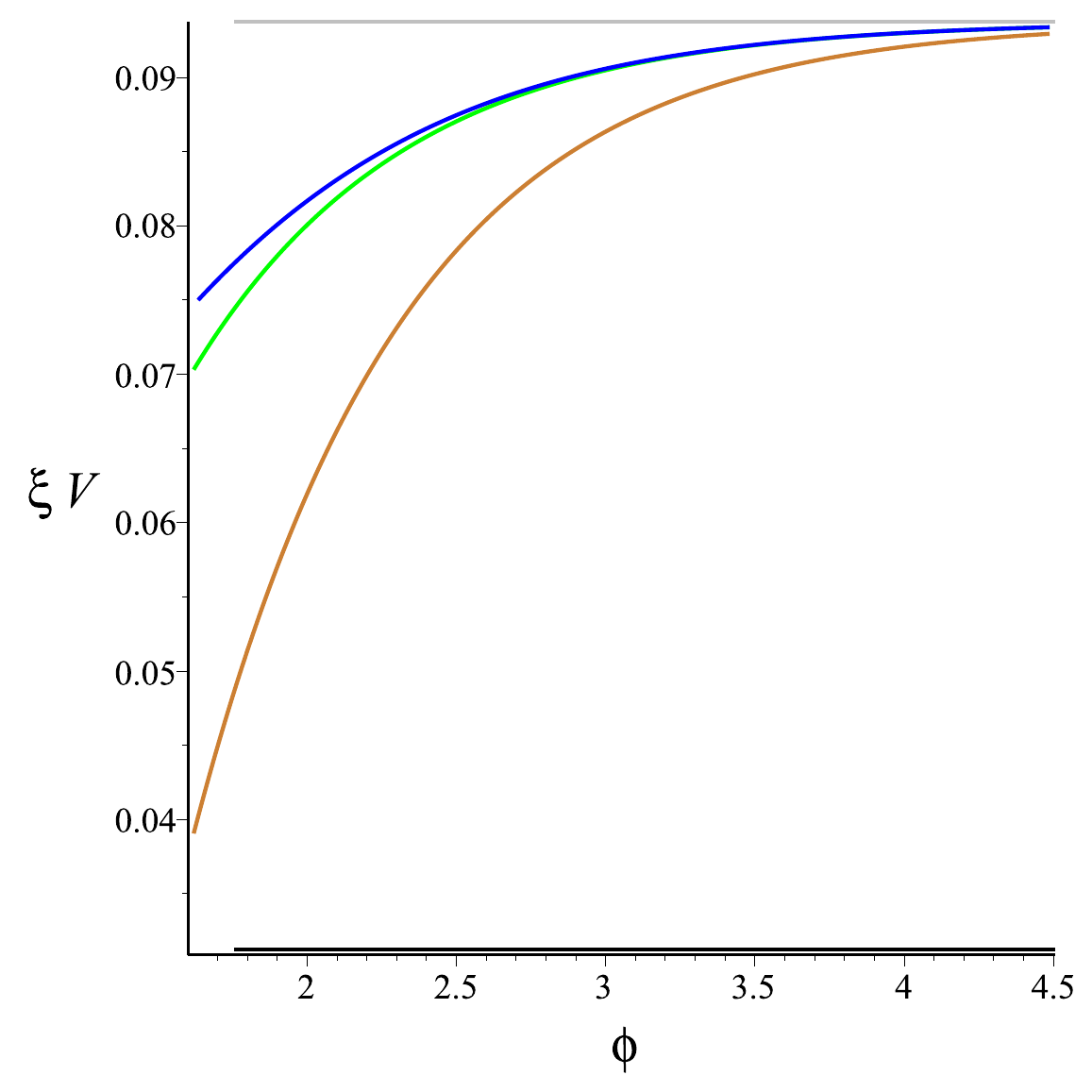}\,
\includegraphics[scale=0.25]{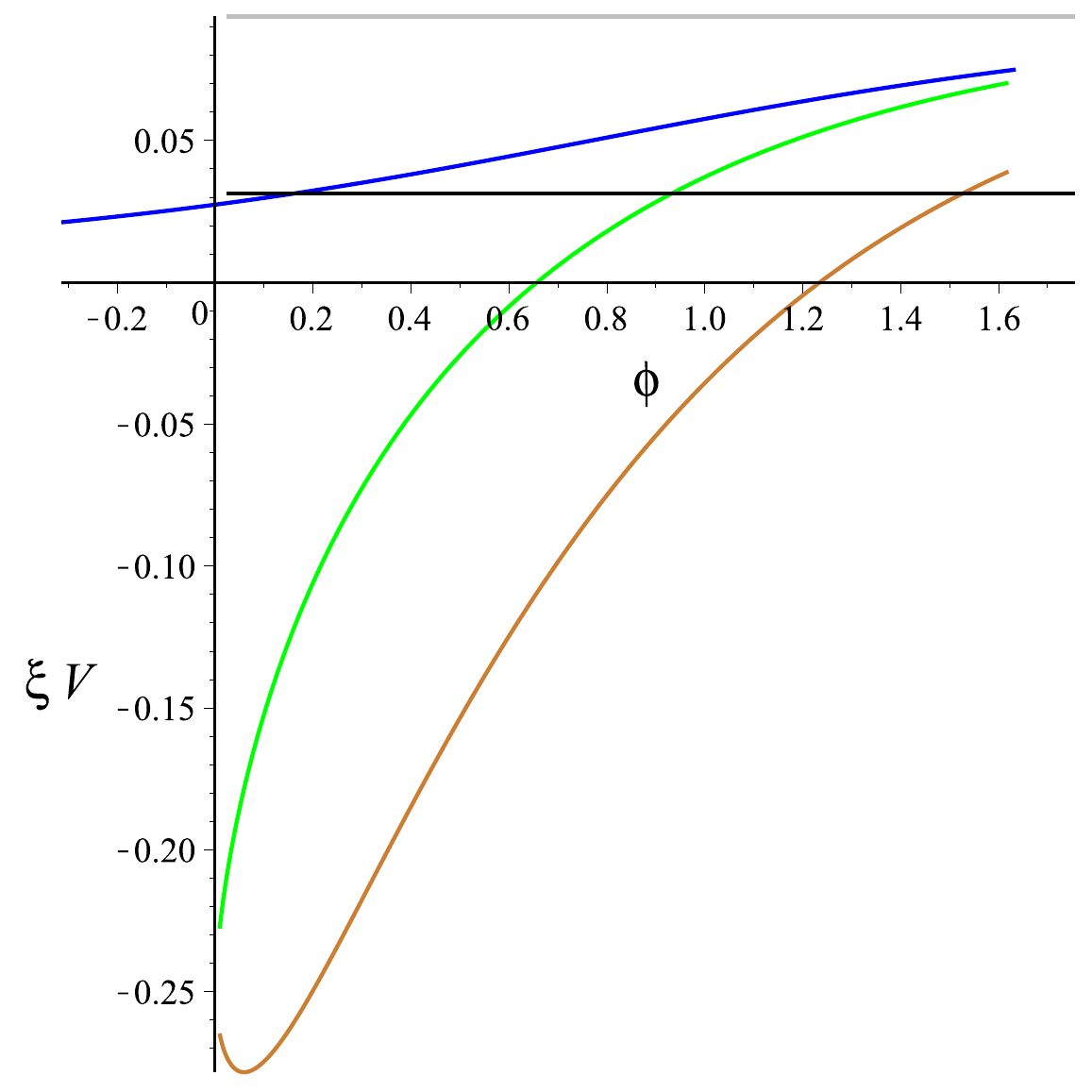}\,
\caption{The behavior of dependence $V\cdot\xi$ on the field $\phi$ (the gray line corresponds to the standard slow-roll approximation, the blue line -- to the approximation \eqref{V1}, the green line -- to the approximation \eqref{V2} ) , the exact considerations (orange line) and $Q\cdot\xi$ (black line) from $N=0..N_b$, from $N=0..7$ and from $N=-0.73..0$ at the following values of the parameters: $c_{sl}=1.7556$, $N_0=1$, $N_b=57.787$, $\xi_0=1.6569\cdot10^{10}/\pi^2$, $Q_0\approx1.8861\cdot10^{-12}\pi^2$, $U_0={M_{Pl}^2}/{2}$, $M_{Pl}=1$.} \label{xiV(phi)}
\end{figure}

\begin{figure}[htp]
\includegraphics[scale=0.25]{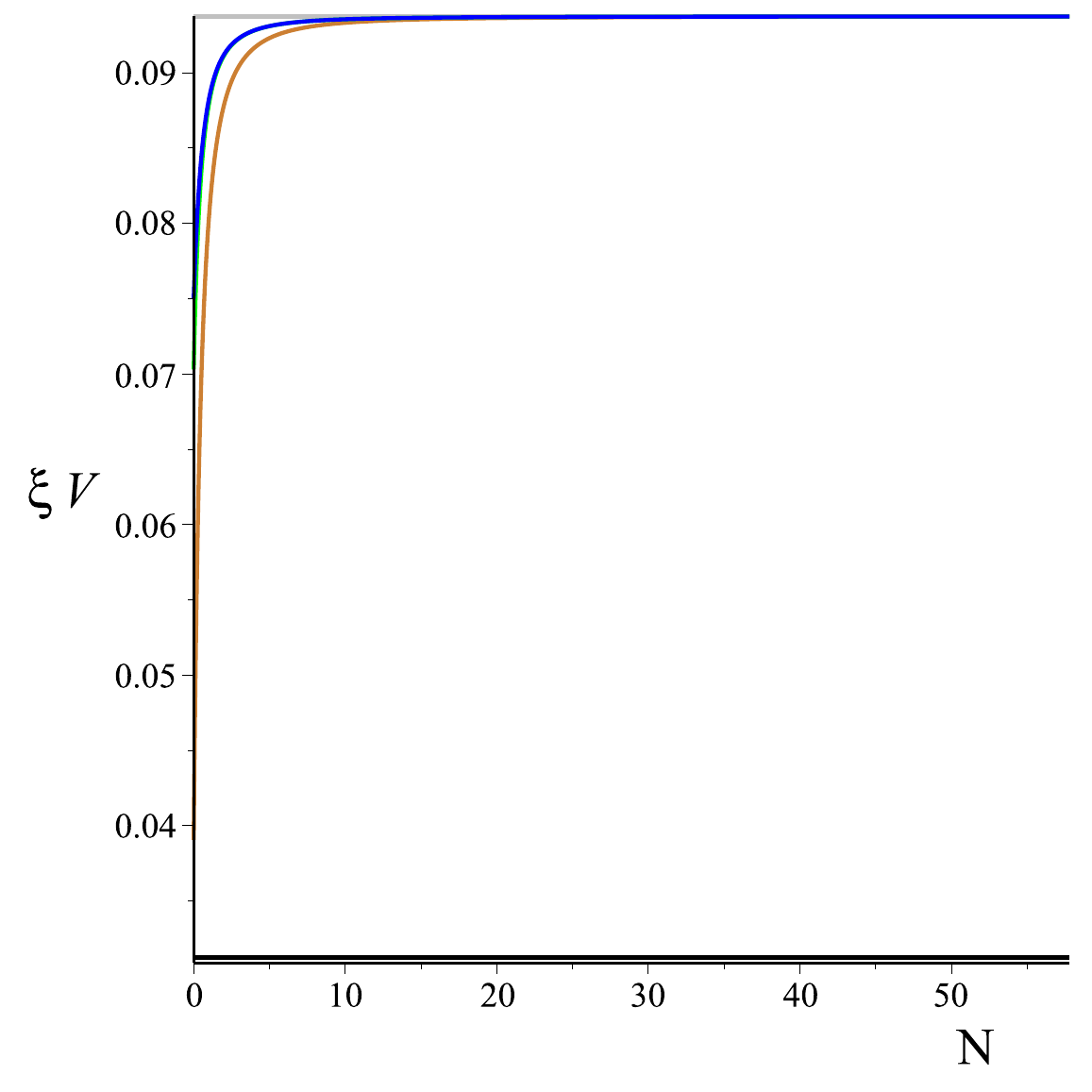}\,
\includegraphics[scale=0.25]{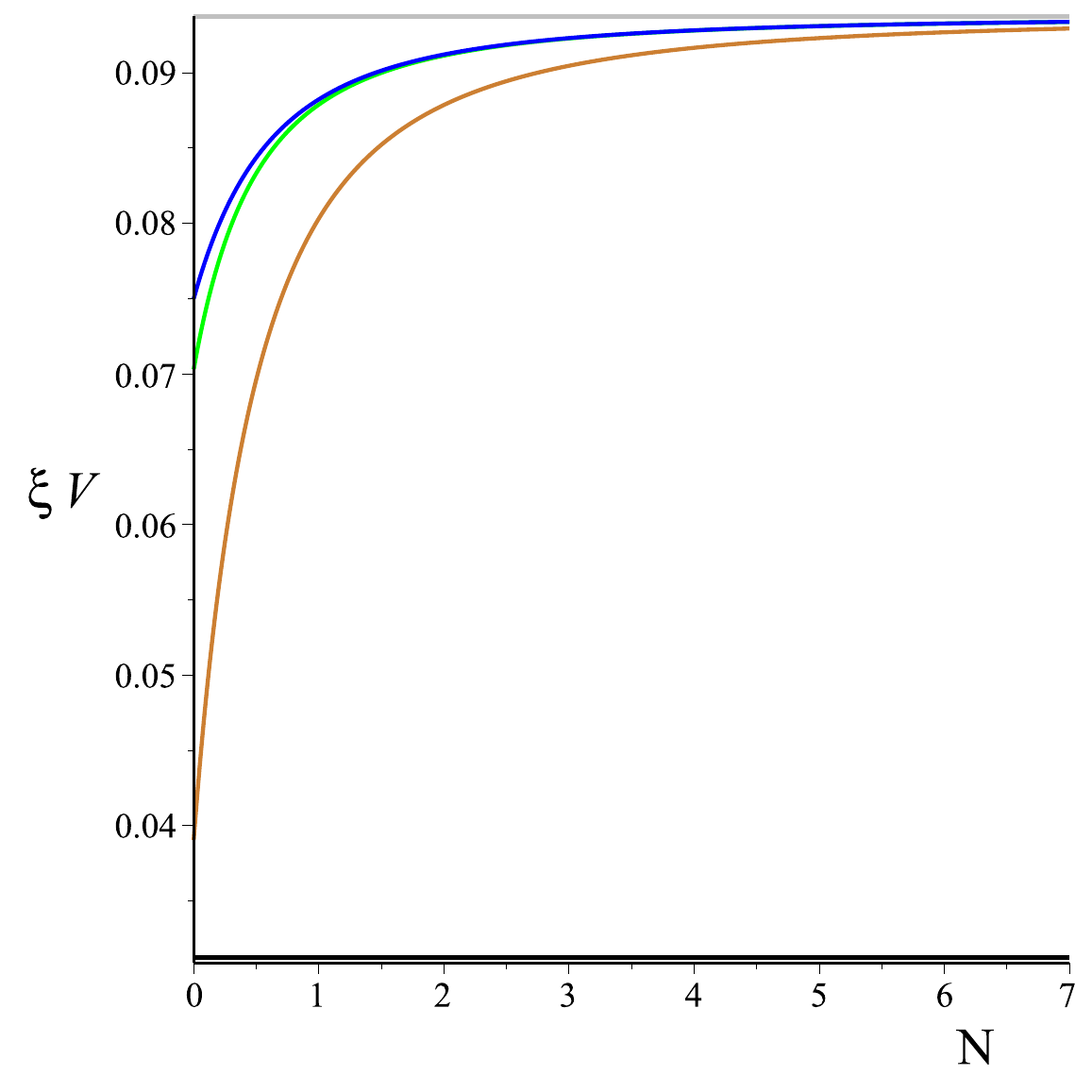}\,
\includegraphics[scale=0.25]{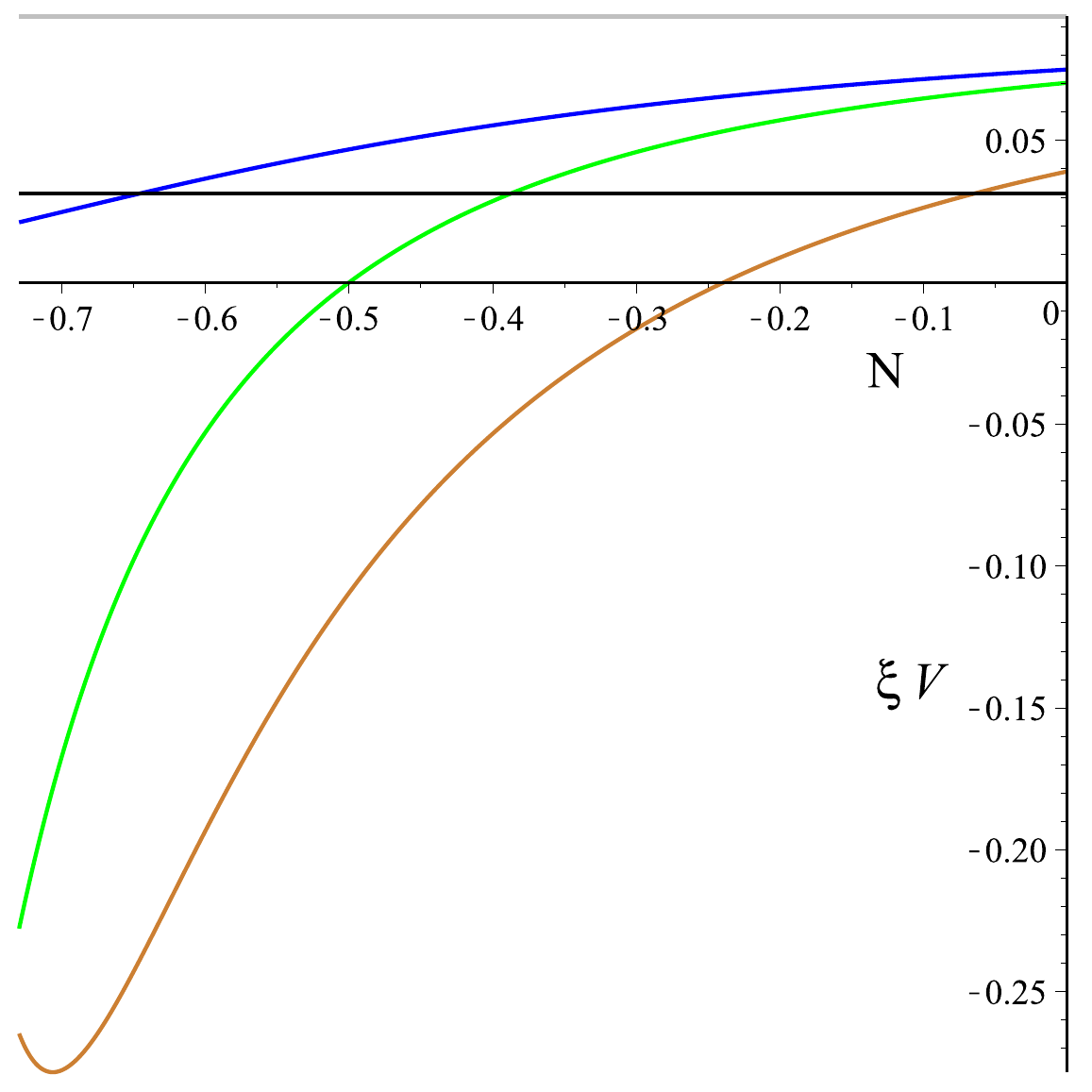}\,
\caption{The behavior of dependence $V\cdot\xi$ (the gray line corresponds to the standard slow-roll approximation, the blue line -- to the approximation \eqref{V1}, the green line -- to the approximation \eqref{V2} ,  the exact considerations -- to orange line)  and $Q\cdot\xi$ (black line)  on the field e-folding number $N$ from $N=0..N_b$, from $N=0..7$ and from $N=-0.73..0$ at the following values of the parameters: $c_{sl}=1.7556$, $N_0=1$, $N_b=57.787$, $\xi_0=1.6569\cdot10^{10}/\pi^2$, $Q_0\approx1.8861\cdot10^{-12}\pi^2$, $U_0={M_{Pl}^2}/{2}$, $M_{Pl}=1$.}\label{xiV(N)}
\end{figure}

\begin{figure}[htp]
\includegraphics[scale=0.25]{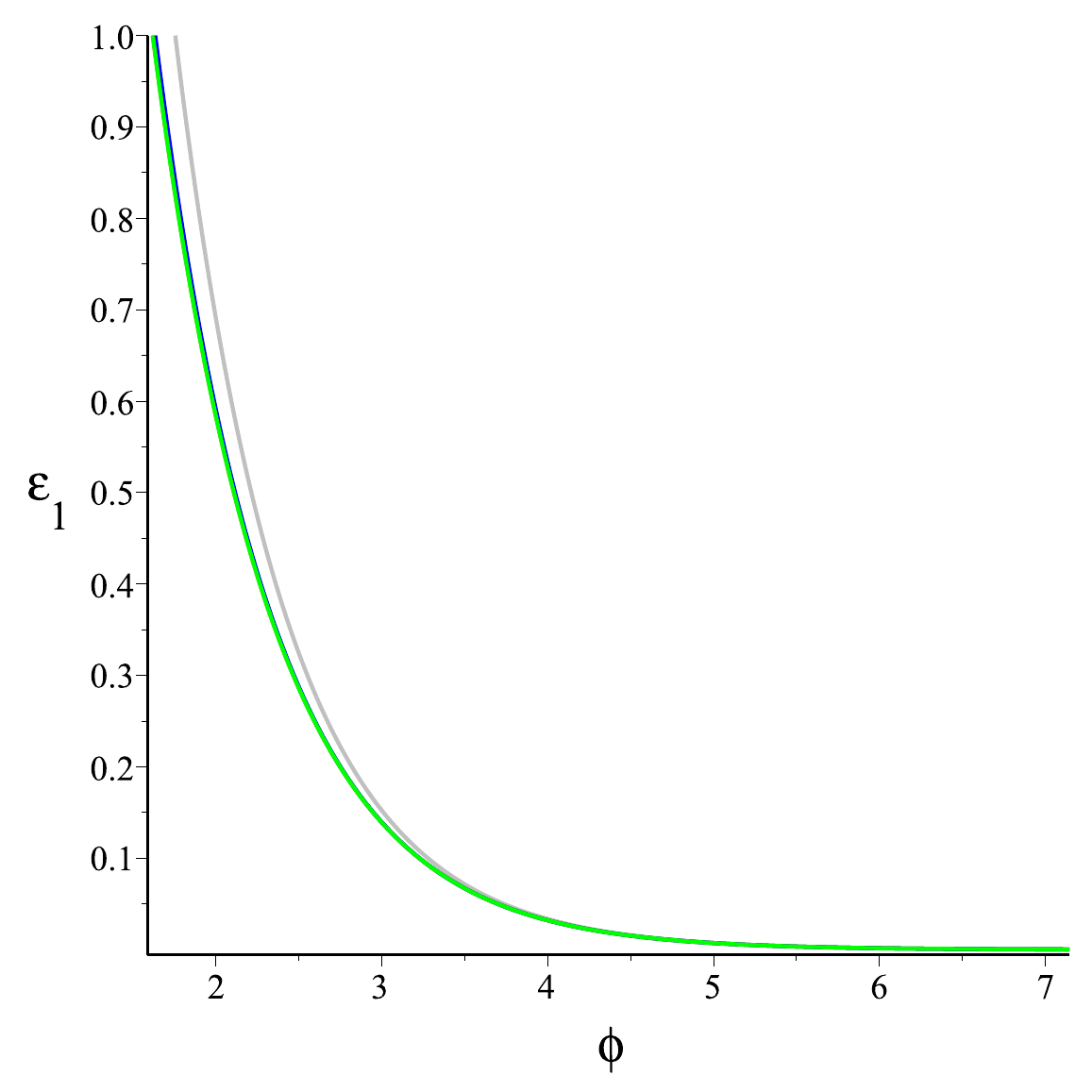}\,
\includegraphics[scale=0.25]{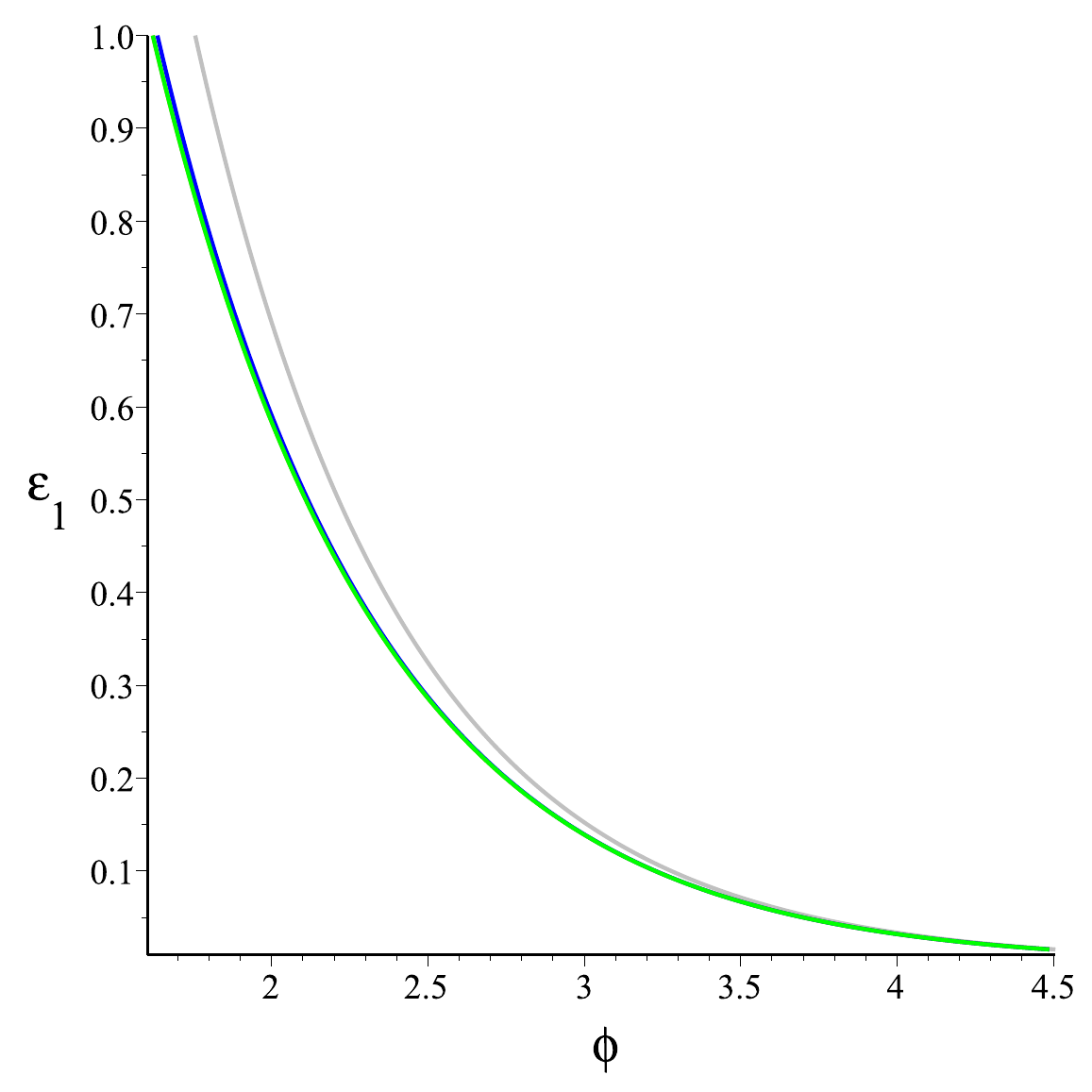}\,
\includegraphics[scale=0.25]{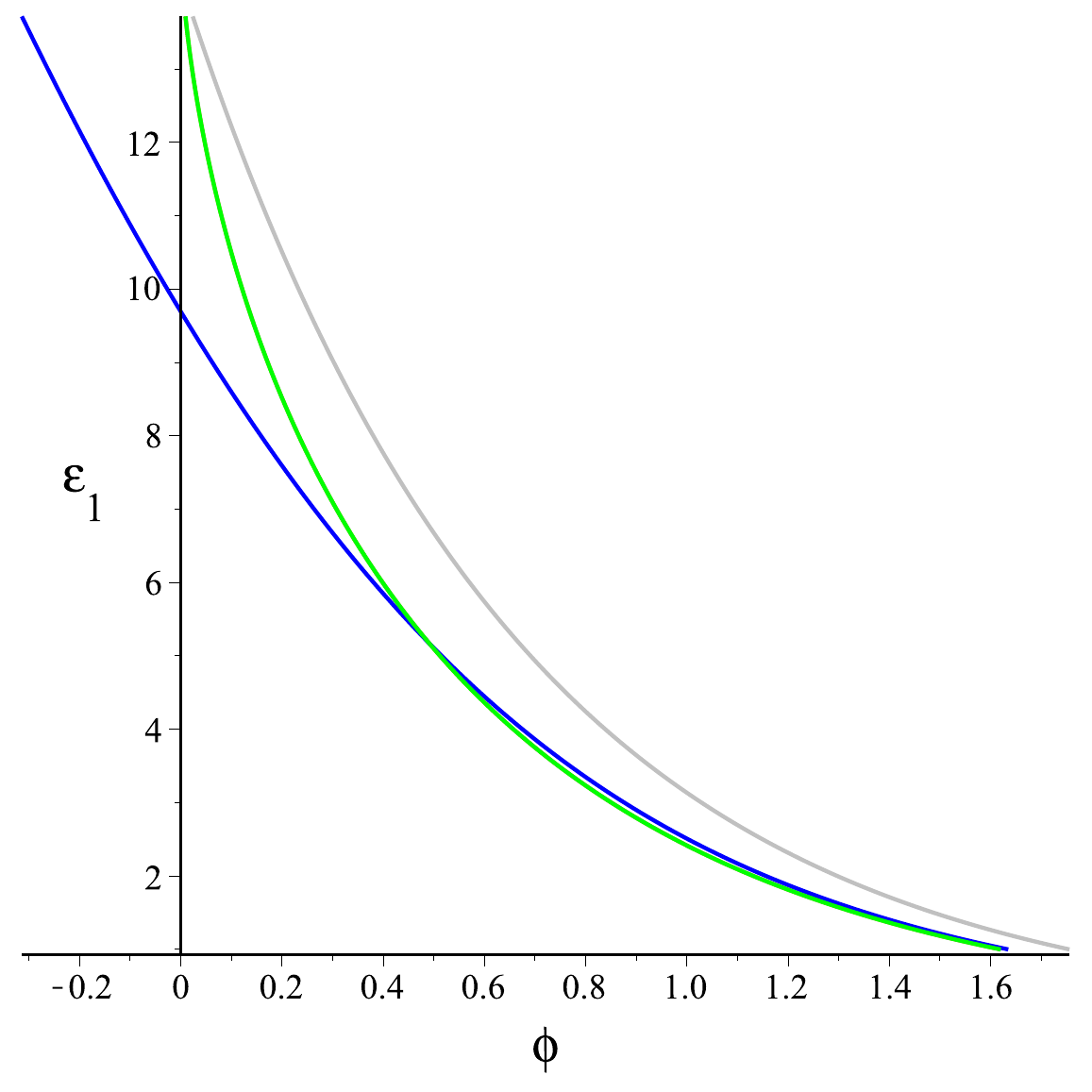}\,
\caption{The behavior $\epsilon_1(\phi)$ (the gray line corresponds to the standard slow-roll approximation, the blue line -- to the approximation \eqref{V1}, the green line -- to the approximation \eqref{V2} and the exact behavior coincides with green line) at the following values of the parameters: $c_{sl}=1.7556$, $N_0=1$, $N_b=57.787$, $\xi_0=1.6569\cdot10^{10}/\pi^2$, $Q_0\approx1.8861\cdot10^{-12}\pi^2$, $U_0={M_{Pl}^2}/{2}$, $M_{Pl}=1$.  The left picture corresponds to the interval $N=0..N_b$ , the central picture corresponds to the interval $N=0..7$ and the right picture corresponds to the interval $N=-0.73..0$ .} \label{dQ(N)}
\end{figure}

We present the different behavior of $\xi(\phi)$ for different types of slow-roll approximation and exact behavior in Fig.\ref{xi(phi)}. At the same time the behavior of $\xi(N)$ does not depend from the type of approximation ($\xi(N)$ is presented in Fig.\ref{xi(phi)}).
In the case of the standard slow-roll approximation, the multiplication $\xi\cdot V$ is a constant during inflation. In the case of the exact expression and the extended slow-roll approximations, the multiplication $\xi\cdot V$ is a constant  near the start of inflation  only. The diviation increases with decreasing numbers of e-folds and value of the field.
The dependence $\xi\cdot V$ on $\phi$ is presented in Fig. \ref{xiV(phi)}.
The dependence $\xi\cdot V$ on $N$ is presented in Fig. \ref{xiV(N)}. The quantity $Q\cdot\xi$ is a constant during inflation.
The exit from inflation is determined by the equality of the first slow-roll parameter to one $\epsilon_1=1$. Within the framework of the considering model, the exit from inflation takes place if $N=0$, regardless of approximations types. If we consider the dependence of the first slow-roll parameter on the field, then the value of field at which the exit from inflation takes place is related with type of considering approximation. The behavior of the first slow-roll parameter on the field is presented in Fig. \ref{dQ(N)}.

\end{document}